\documentclass[11pt]{article}

\usepackage{sectsty}
\usepackage{graphicx}

\newif\ifextra
\extrafalse

\usepackage[a4paper,margin=1in]{geometry}

\title{Functional Renormalization for Elastic Burgulence}
\author{Johannes Conrad\\  Martin Oberlack \\
\small \texttt{conrad@fdy.tu-darmstadt.de}}

\date{\today}

\usepackage{mathtools} 
\usepackage{amssymb}   
\usepackage{siunitx}   
\usepackage{amsmath} 
\usepackage{bm}
\usepackage{tikz}
\usepackage{amsthm}

\usepackage{algorithm}
\usepackage{algpseudocode}

\usepackage{tikz-feynman}
\usetikzlibrary{positioning,shapes.misc}

\newcommand{\Lie}[1]{L_{#1}}

\newcommand{\ccd}[1]{\overset{\circ}{#1}}
\newcommand\identity{1\kern-0.25em\text{l}}
\newcommand{\tr}[1]{\mathsf{tr}\left(#1\right)}
\newcommand{\D}{\mathcal{D}}
\newcommand{\pr}{\mathsf{pr}\,}
\newcommand{\EL}{\mathsf{E}}
\newcommand{\Fd}[1]{\mathsf{D}_{#1}}
\newcommand{\ed}{\tilde{d}}

\DeclareFontFamily{U}{wncy}{}
\DeclareFontShape{U}{wncy}{m}{n}{<->wncyr10}{}
\DeclareSymbolFont{mcy}{U}{wncy}{m}{n}
\DeclareMathSymbol{\Sh}{\mathord}{mcy}{"58}

\newcommand{\const}{\small\text{const.}}

\usepackage[mathscr]{euscript}

\DeclareSymbolFont{rsfs}{U}{rsfs}{m}{n}
\DeclareSymbolFontAlphabet{\mathscrsfs}{rsfs}

\usepackage[style=numeric,sorting=nyt,backend=biber,defernumbers,doi=true,isbn=false,url=false,eprint=false,giveninits=true,maxnames=10]{biblatex}
\usepackage{xpatch}
\begin{document}
\maketitle	
\begin{abstract}
    We formulate elastic and elasto-inertial turbulence in the Martin-Siggia-Rose path-integral formalism and develop a systematic source-extended symmetry algorithm to derive Ward identities directly from the Euler-Lagrange equations. These identities provide nonperturbative constraints and a principled foundation for constructing closure schemes. 
As a dimensionally reduced model for elastic turbulence, we propose an extended Burgers equation that preserves the characteristic coupling between the extra stress and velocity gradient, while remaining simple enough for first controlled calculations. In particular, we obtain an extended set of Ward identities that strongly constrains admissible closures and provides insight into the scaling behaviour near the fixed point.
\end{abstract}

\section{State of the art}
\textbf{Elastic and Elasto-Inertial Turbulence.}
The transition from a smooth, i.e., laminar flow regime, to a disordered turbulent state is associated with classical Newtonian fluids with a simple linear constitutive relation between the strain rate tensor $\bm{D}=\nabla \bm{u} + (\nabla \bm{u})^T$ and the Cauchy stress tensor $\bm{\tau} \propto \bm{D}$.
The transition is thus triggered by instabilities, in which disturbances are no longer dampened by the inherent viscosity, and a turbulent state is maintained by the modal coupling of the nonlinear advective term driving a cascade of energy across scales.
However, if the viscous material properties are extended by elastic properties, for example, in polymer solutions, the transition can be triggered significantly earlier.
Viscoelastic fluids introduce an additional dynamical variable in the form of the polymer conformation or elastic stress tensor, which couples nonlinearly to the velocity gradient.
The classical Navier-Stokes system needs to be altered, and a transport equation for the viscoelastic stress tensor is coupled in, whereby we will subsequently focus on the Oldroyd-B model \cite{Oldroyd1950}. With the time $t >0$ and  the space variable $\bm{x}\in D \subseteq \mathbb{R}^d$, we search for the velocity field $\bm{u} =\bm{u}(t,\bm{x})  \in \mathbb{R}^d$, the stress tensor $\bm{\sigma} =\bm{\sigma}(t,\bm{x})  \in \mathbb{R}^{d \times d}$ and the scalar pressure field
$p=p(t,\bm{x})$  that  satisfy
\begin{align}
		\partial_t \bm{u}  + \nabla \cdot (\bm{u} \otimes \bm{u}) + \nabla \pi &= \nu \nabla \cdot \bm{\sigma} + \nu \beta \Delta \bm{u} + \bm{\xi} \label{Imp}\\
		\nabla \cdot \bm{u} &= 0 \\
		\lambda \Lie{\bm{u}} \bm{\sigma}&= - \bm{\sigma} + \nu (1-\beta)\left(\nabla \bm{u}+ (\nabla \bm{u})^T\right).   \label{Cont}
\end{align}
In \eqref{Imp}, $\nu$ denotes the total viscosity, the inverse Reynolds number in dimensionless variables $Re= U L / \nu$. The Cauchy stress tensor is thus defined as $\bm{\tau}= \nu (\bm{\sigma} + \beta \bm{D})$. $\beta \in [0,1]$ denotes the ratio of the solvent viscosity to the total viscosity, and $\lambda$ denotes the polymer relaxation time, i.e., the Weissenberg number in dimensionless units $W=\lambda / T$. The latter encodes the ratio of the relaxation time $\lambda$, the intrinsic timescale of the non-Newtonian solvent, and the bulk flow timescale $T$. $\bm{\xi}$ denotes a solenoidal random forcing. 
Finally, we introduce the material frame indifferent \cite{Marsden2015} lower convective derivative
\begin{align}
    \Lie{\bm{u}} \bm{\sigma} &= D_t \bm{\sigma} + \nabla \bm{u} \cdot \bm{\sigma} + \bm{\sigma} \cdot \left(\nabla \bm{u} \right)^T, \\
    D_t  &= \partial_t + \bm{u} \cdot \nabla,
\end{align}
which non-linearly couples the velocity gradient to the stress tensor $\bm{\sigma}$. 
This coupling generates nonlinear dynamics even in the absence of inertial advection and may itself sustain a turbulent state, which was first experimentally observed by Groisman and Steinberg \cite{Groisman2000}, who showed that dilute polymer solutions exhibit a spatially disordered flow state at vanishing Reynolds number once elastic stresses are sufficiently amplified. A turbulent regime, driven solely by the nonlinear interaction of the stress tensor with the flow field, that is in the Stokesian limit $\nu \to \infty$, is commonly dubbed elastic turbulence (ET). In contrast, the viscoelastic regime, also dubbed elasto-inertial turbulence (EIT), exhibits both a relevant nonlinearity stemming from the constitutive equation \eqref{Cont} as well as the advection term of the momentum equation \eqref{Imp}. 
Subsequent experiments established the robustness and generality of this phenomenon across curvilinear flows \cite{Groisman2004}. 
Numerical studies have demonstrated that ET arises in homogeneous and spatially periodic flows. For instance, Berti et al.~\cite{Berti} showed that ET emerges in two-dimensional Kolmogorov flows of an Oldroyd-B fluid, establishing that ET need not be driven by boundary effects. This result motivates the study of homogeneous and isotropic turbulence (HIT) in viscoelastic fluids, the statistical state that is best understood in the Navier-Stokes setting.
Statistical analyses of EIT have shown that polymer elasticity significantly modifies the scale-by-scale rate of energy transfer. In contrast to Navier-Stokes turbulence, where the energy balance is based on the kinetic energy only, viscoelasticity leads to an additional contribution in the form of a polymeric energy term, encoding the energy stored as a deformation of the polymer microstructure $E=E_{k}(|\bm{p}|) + E_{p}(|\bm{p}|)$, where $\bm{p}$ denotes the spatial wavevector. Indices $k$ and $p$ denote the kinetic and polymeric contribution respectively.
Within a turbulent subrange characterized by a constant rate of energy dissipation $\langle\varepsilon \rangle = \langle \bm{\tau}: \bm{D} \rangle$, the spectral energy densities are expected to scale as power laws. Numerical studies of HIT \cite{Rosti2023} suggest $E_p \sim |\bm{p}|^{-1.35}$ and $E_k \sim |\bm{p}|^{-2.3}$, in a limited range for the parameter $\lambda$. Notably, the kinetic energy scaling differs significantly from Kolmogorov’s theory of inertial turbulence \eqref{EKol}. Furthermore, there are currently no analytical predictions for these scaling exponents.
Recent work has emphasized the intermittent and small-scale nature of ET. Strong intermittency and non-smooth velocity fields have been reported \cite{Singh2024}, while unifying perspectives suggest that ET and EIT represent asymptotic limits of a single viscoelastic turbulent dynamical state \cite{FoggiRota2024}. Recent studies show that ET-like dynamics persist at scales below the Kolmogorov scale in fully developed inertial turbulence \cite{Garg2025}, accompanied by localized elastic stress structures in homogeneous flows \cite{Watanabe2025}.
From a theoretical perspective, ET provides an extension of the familiar notion of inertial turbulence beyond the Navier-Stokes case. Nonetheless, only very limited insights into elastic and elasto-inertial turbulence have thus far been derived from a theoretical standpoint.
\\
\textbf{Statistical Methods in Turbulence.} \\
Throughout the past century, three distinct but equivalent formulations of statistical hydrodynamics have been developed to describe Navier-Stokes turbulence. The predominant description in engineering applications is the use of statistical moments as forwarded by Osborne Reynolds \cite{Reynolds1895}. By decomposing the flow into its mean $\left\langle \bm{u}\right\rangle$ and fluctuation $\bm{u}'$, the famous Reynolds Averaged Navier-Stokes equations emerge, wherein the closure problem of turbulence emerges as the Reynolds stress tensor $\left\langle \bm{u}' \otimes \bm{u}' \right\rangle$, which can not be determined from first order statistical moments. 
Through the introduction of infinitely many higher-order multi-point moments, the formally closed Friedmann Keller hierarchy of partial differential equations arises \cite{Oberlack2010}.
Although the Reynolds averaged Navier-Stokes and moment equations have seen wide adoption in engineering applications through truncation and the application of closure models, the increasing number of unclosed terms at each level of the hierarchy often makes the treatment of high-order equations untractable.
Whereas the description of turbulence in terms of statistical moments exhibits an increasing number of unclosed terms at each level, this is not the case when considering the statistical description of turbulence in terms of multi-point probability density functions as teh econd approach. Within the Lundgren \cite{Lundgren1967}, Monin \cite{Monin1967}, Novikov \cite{Novikov1967} hierarchy of probability density functions, only a single unclosed variable emerges at every level. 
The third, and equivalent, statistical description of turbulence is formulated in terms of the characteristic functional, whose dynamics is governed by the Navier-Hopf functional differential equation \cite{Hopf1952}.
Although all three of the above methods are formally closed, one must still perform a truncation of the hierarchical structure or approximation of the functional differential equation to perform calculations. Unfortunately, all presently available closure schemes are heavily based upon heuristic arguments and are unsatisfactory in their predictive quality.
Other methods of describing turbulence are usually based upon a combination of heuristic reasoning and simple dimensional analysis, such as the log-law of the wall or Kolmogorov's description of the inertial subrange.
In the statistically steady regime and with respect to a probability measure on the space of velocity fields, Kolmogorov postulates a constant flux of kinetic energy from large to small scales, resulting in the energy cascade
\begin{align} \label{EKol}
    E(|\bm{p}|) = C_K \langle \varepsilon \rangle^{2/3} |\bm{p}|^{-5/3}, \quad \varepsilon = 2 \nu \bm{D}: \bm{D},
\end{align}
with $\varepsilon$ the rate of energy dissipation, $\bm{p}$ the spatial wavenumber, and $C_K$ the a priori undetermined Kolmogorov constant. Therein, $\langle\cdot\rangle$ denotes the mean with respect to the probability measure. This scaling supposedly holds for $\eta \ll 1/|\bm{k}| \ll \ell$ with $\ell$ and $\eta$ being the integral and the viscosity-dependent Kolmogorov length scales. 
The subrange postulate leads to a relation of statistical moments of velocity increments (structure functions) to the mean rate of energy dissipation. In the inertial subrange, an $m^\mathrm{th}$ order structure function satisifies
\begin{equation}\label{KolmogorovStruc}
   \left\langle | \left(\bm{u}(\bm{x}+\bm{r})-\bm{u}(\bm{x}) \right) \cdot (\bm{r} / |\bm{r}| )|^m \right\rangle \sim |\bm{r}|^{m/3} \langle \varepsilon \rangle^{m/3} \left(\frac{|\bm{r}|}{L}\right)^{\zeta_m-m/3}.
\end{equation}
The vector $\bm{r}$ fixes the direction of correlation, and the correlation distance $r=|\bm{r}|$ satisfies $\eta \ll r \ll \ell$. This implies the validity of \eqref{KolmogorovStruc} in the low viscosity regime when nonlinear hyperbolic transport prevails.   
Dimensional analysis yields $\zeta_m=m/3$, while only for $m=3$, an exact result can be deduced from the von  K\'arm\'an-Howarth equation \cite{Pope2015}. Generally, \eqref{KolmogorovStruc} seems to be valid only for $m=2,3$, while measurements indicate significant deviations for larger $m$.
To go beyond classical hierarchical closures, it is thus useful to adopt a path-integral formulation reminiscent of quantum field theory, with the aim of applying approximation methods that have been developed in this well-studied field.
\\
\textbf{Path Integral Formulations for Non-Equilibrium Statistical Mechanics.} 
Conceptually, the path integral formulation is a direct construction of a functional probability measure which can be used to compute the mean of arbitrary observables by summing over all possible field configurations that the dynamical variables e.g~the velocity $\bm{u}$, pressure $p$, and stress $\bm{\sigma}$ may possess, weighted by their respective probability.
Describing the dynamics of stochastic differential equations via path integrals has a long history. The first functional integral representations for stochastically forced equations were derived for simple Langevin equations by Onsager and Machlup \cite{Onsager1953}. However, these early formulations were restricted to simple Langevin equations with additive noise and could not easily handle nonlinear interactions or ensure a causal response of the dynamical fields to the noise. 

These problems are avoided by the method of Martin Siggia and Rose (MSR) \cite{Martin1973}, and its extension by Janssen \cite{Janssen1976} and Dominicis \cite{DeDominicis1976}, whose approach overcomes these limitations of the Onsager-Machlup formalism by introducing response fields. These auxiliary fields are dual to the dynamical variables and enforce the stochastic dynamics, we denote them as $\overline{\bm{u}}$ and similarly for the other variables in the following. By choosing a discretization scheme of the underpinning stochastic differential equation (SDE), usually Ito for simplicity of the involved functional Jacobian \cite{Canet2011}, one can derive a local action that encodes the dynamics of the underpinning SDE.
%
Given the governing equations \eqref{Cont} with Gaussian zero-mean forcing $\bm{\xi}$ correlated as $\langle \bm{\xi}(x,t) \bm{\xi}(x',t') \rangle = \bm{Q}(x-x') \delta(t-t')$ and using the MSR formalism, the moment generating functional can be written as
\begin{align} \label{genfun}
    \mathcal{Z}[J,\overline{J}] = \left\langle \exp \left( \int_{x,t} \; \bm{J} \cdot \bm{\phi} + \overline{\bm{J}} \cdot \overline{\bm{\phi}}\right) \right\rangle = \iint \D \bm{\phi}  \D [i \overline{\bm{\phi}} ]   \exp \left(- S +\int_{x,t} \; \bm{J} \cdot \bm{\phi} + \overline{\bm{J}} \cdot \overline{\bm{\phi}} \right), 
\end{align}
where $\bm{\phi}=(\bm{u},\bm{\sigma},p)$ denote the original fields, $\overline{\bm{\phi}}=(\overline{\bm{u}},\overline{\bm{\sigma}},\overline{p})$ are the response fields, and $\D$ denotes a functional measure. The functional integral with respect to $\D \bm{\phi}  \D [i \overline{\bm{\phi}} ]$ then conceptually encodes the sum over all possible field configurations. The action $S[\bm{\phi},\overline{\bm{\phi}}]$ is defined as follows
\begin{align} \label{SOB}
    S &= \int_{x,t} \overline{\bm{u}} \cdot \left(\nu^{-1} \, D_t \bm{u} - \nabla \pi - \nabla \cdot \bm{\sigma} - \beta \Delta \bm{u} \right) +\overline{\pi} \nabla \cdot \bm{u} 
    +  \overline{\bm{\sigma}} : \left(\lambda \Lie{\bm{u}}\bm{\sigma} + g \bm{\sigma} - (1-\beta)  \bm{D} \right) \\
    &+ \int_{x',x,t} \overline{\bm{u}}(x,t) \cdot \bm{Q}(|x-x'|)  \cdot \overline{\bm{u}}(x',t) \nonumber,
\end{align}
with the first three terms representing the deterministic dynamics. For example, the response pressure $\overline{p}$  enforces incompressibility by acting as a Lagrange multiplier. The last term in \eqref{SOB} then encodes the stochastic forcing.  
Unfortunately, infinite-dimensional path integrals such as \eqref{genfun} are only solvable in the rarest of circumstances, for instance, if the underlying theory is free (Gaussian) or for certain two-dimensional conformally invariant models \cite{Peskin2018, Klauder2011, DiFrancesco1997}. Thus, one must resort to perturbative or non-perturbative approximations. Furthermore, the probability measure that is represented by the path integral \eqref{genfun} may not even be well defined as is \cite{Klauder2011}.
To overcome these difficulties, we aim to apply the renormalization group (RG) to the field theory governing ET \eqref{SOB}, which was first developed as a tool to study how the behavior of stochastic systems, in quantum field theory and non-equilibrium statistical mechanics, changes across scales. 
In Navier-Stokes turbulence, kinetic energy is transferred through a cascade of vortices from large to small scales, and one observes that the rate of energy transfer across wavenumber shells is universal within the scale-invariant inertial subrange. 
Turbulent dynamics in this subrange can thus be thought of as a non-Gaussian fixed point in the language of field theory, which can be probed using RG methods.
The Wilsonian, perturbative RG approach originated in the study of high-energy particle physics to explain the emergence of ultraviolet divergences and how they can be absorbed into scale-dependent couplings, that is, parameters such as the electron charge. The RG has also proven indispensable in non-equilibrium statistical mechanics, where it can be used to obtain an accurate description of phase changes and critical exponents. 
It was first cast into a computationally substantive form by Kenneth Wilson \cite{Wilson1971}, whom it is commonly named after. The underpinning idea of which is to average over fluctuations at small scales, that is, large wavenumbers $|\bm{p}|$ or momenta in the language of quantum mechanics. This process then modifies the effective dynamics of large-scale modes, and can be done in iterative steps. This introduces scale-dependent effective material parameters in the process. 
Returning to the setting of Navier–Stokes turbulence, such a parameter is the effective viscosity at momentum, or wavenumber, scale $k$
$\nu_{k} \equiv \hat{\nu} k^{-\Delta_\nu}$ . Here, $\Delta_\nu$ denotes the total scaling dimension made up of an anomalous and canonical contribution, the former approaches a constant value at the scale-invariant fixed point. $\hat{\nu}(k)$ is the corresponding dimensionless coupling.
As it turns out, the assumption of a scale-invariant energy injection rate and Galilean invariance already implies $\nu_{k} = \varepsilon^{1/3} k^{-4/3}$ in this particular case.
The application of the RG to Navier-Stokes turbulence was first proposed by Forster, Stephen, and Nelson \cite{Forster1977} and further refined by Yakhot and Orszag \cite{Yakhot1986}. While these approaches were pioneering, they relied on a perturbative expansion in the Reynolds number, which is not valid in any realistic turbulent regime.
Mathematically, their perturbative approach in the spirit of Wilsonian RG necessitated the introduction of an artificial power-law forcing to obtain a scale-invariant inertial subrange. This approach cannot describe a finite energy injection rate and violates the separation of turbulent scales. Additionally, it has been observed that RG-based turbulence models are not able to accurately depict physical effects, such as streamline curvature.

A decisive breakthrough came with the development of the functional renormalization group (fRG) governed by the Wetterich flow equation \cite{Wetterich1993}. Unlike perturbative RG, the fRG tracks the full effective action $\Gamma_{k}$, which contains the same information as $W_k=\ln \mathcal{Z}_{k}$, as a function of scale $k$. This allows for nonperturbative truncations and physically realistic forcing profiles.
This formalism allows for the preservation of the actions symmetries across scales, enabling accurate closure schemes and reliable predictions. The construction of closure schemes in this setting is well researched and highly robust. That is, even a relatively crude approximation for the effective action, such as the leading potential approximation, can lead to astonishingly accurate results. 
\\
\textbf{Application of the fRG to Non-Equilibrium Statistical Mechanics and Navier-Stokes Turbulence.} 
The fRG has been applied successfully to a number of statistical field theories, from relatively simple theories describing equilibrium critical phenomena such as the $\phi$-four theory and the $O(N)$ model \cite{Blaizot2006, Blaizot2006a, Benitez2008}, where the fRG provides quantitatively accurate critical exponents and phase diagrams \cite{Tetradis1994}, and to quantum field theory. Beyond equilibrium, the methods developed in the field have proven equally powerful for problems formulated in the MSR path integral, such as \eqref{genfun} with action \eqref{SOB} for \eqref{Cont}.
A paradigmatic example is the Kardar-Parisi-Zhang equation \cite{Kardar1986}, modeling kinematic roughening and stochastic surface growth. Here, the fRG and wavenumber resolving closure schemes, such as a symmetry-based Vertex function expansion \cite{Canet2010} or first-order closure \cite{Canet2011a}, were able to capture the strong coupling fixed point inaccessible through perturbative methods and accurately reproduced the exact solution available in $1+1$-dimensions \cite{Sasamoto2010}.  
The fRG has also been applied to Navier-Stokes turbulence in two and three dimensions with great success \cite{Canet2016}. Using wavnumber resolving approximations, critical exponents and the Kolmogorov energy cascade $E(|\bm{p}|) = C_K \varepsilon^{2/3} |\bm{p}|^{-5/3}$ could be derived in three-dimensions at the RG fixed point realized at $k \to 0$. 
Notably, the fRG reveals not only the qualitative behaviour $E \sim |\bm{p}|^{-5/3}$, which Kolmogorov obtained from dimensional reasoning, but can be used to calculate the quantitative value of the Komogorov constant $C_K \approx 1.572$, which slightly depends on the forcing's shape encoded through the correlation tensor $\bm{Q}^{-1}$. This value agrees with experimental data $C_K \approx 1.58$ \cite{Donzis2010} and differs significantly from the perturbative result of $C_K=1.617$ obtained by Yakhot and Orszag \cite{Yakhot1986}.
Additionally, the Kolmogorov $4/5$'th law $\langle ( \bm{u}(\bm{x}+\bm{r})-\bm{u}(\bm{x}) )\cdot \bm{r} )^3 \rangle=- 4/5 \varepsilon |\bm{r}|$ for the longitudinal structure function, valid in the inertial subrange, could be derived through the fRG \cite{Canet2022}. 
Furthermore, the remarkable observation was made that the high-frequency $|\bm{p}| \gg k$ vertex functions, that is, the connected correlation functions, are determined to second order in $\bm{p}$ by exact Ward identities. Therein, $k$ is the wavenumber scale at which forcing occurs. This observation can be used to derive a diagrammatic closure of the Wetterich flow equation. 
Due to analyticity of the Wetterich equation, this expansion becomes exact for $|\bm{p}| \to \infty$, that is, at small scales, and constitutes one of only a handful of analytical results in Navier-Stokes turbulence. The expansion accurately describes both short- and long-time correlations analytically. 
In the treatment of two-dimensional turbulence \cite{Tarpin2019}, similar relations were derived and, in the absence of intermittency, give rise to Kraichnan's logarithmic sub-leading correction of the energy cascade $E(|\bm{p}|) \sim |\bm{p}|^{-3} \ln(|\bm{p}| L)$, with $L$ the integral length scale. 
For a review of recent results, we refer to \cite{Canet2022}, whereas a self-contained introduction to the fRG can be found in \cite{Canet2011}, \cite{Dupuis2021}, or \cite{Tetradis1994}.
\\
The application of the FRG to elastic turbulence is more intricate when compared to calssical Navier Stokes turublence due to several different properties of the governing equations.
First and foremost, the high dimensionality of the dynamical variables, the stress tensor $\bm{\sigma}$ in addition to the fluid velocity $\bm{u}$. This renders a closure scheme alike the Blaizot Mendez Weschbor closure intractable. On the other hand, the underlying equations symmetries and Ward Identities do not permit access to the zero-momentum sector as in the Navier-Stokes setting. Whence, there is significantly less structure to base closures upon. 
Finally, since there is no galilean or shift symmetry in the extra stress, the assumption of a vanishing mean stress $\langle \bm{\sigma} \rangle$ is generally false. This leads to the possible emergence of dynamical instabilities of the mean field configuration.
The present work aims to apply the methodology of the fRG to elastic Burgulence, as a dimensionally reduced model problem capable of depicting some, if not all, difficulties that arise in the application to ET and EIT. 
\section{Preliminaries}
Before we delve into the application of the functional renormalization group, let us briefly discuss what it is we wish to calculate. That is the moment generating functional $\mathcal{Z}$. Let $\xi(t)$ be a stochastic variable with probability measure $d \mu = f[\xi(t)] \D \xi$, $f[\xi(t)]$ being the probability density function(al) thereof. The ensemble average of some observable $\mathcal{O(\xi)}$ can then be written as
	\begin{align}
		\langle \mathcal{O} \rangle = \int d \mu(\xi) \mathcal{O}(\xi) = \int \D \xi f[\xi] \mathcal{O},
	\end{align}
    with $f \D \xi$ the probability measure on paths.
	Now, let us assume $\mathcal{O}=\exp(\int dt \;\xi(t) J(t))$, defining $\langle \mathcal{O} \rangle=\mathcal{Z}$ is the functional Fourier transform of the measure $\mu$. We may then take functional derivatives thereof, and evaluate at $J=0$, to obtain statistical moments 
	\begin{align}
		\frac{\delta \mathcal{Z}}{\delta J(s)} \Big|_{J=0} &= \left( \frac{\delta}{\delta J(s)} \int d \mu \; \exp\left(\int dt \;\xi(t) J(t)\right) \right) \Big|_{J=0}= \int d \mu \; \xi(s) .
	\end{align}
	wherein the functional derivative can be introduced through a limit as follows. Let $\delta_\epsilon(t)$ be a function with bounded support in the open ball $|t|<\epsilon$ and normalized such that
    \begin{align}
    	\int dt \;\delta_\epsilon(t)=1,
    \end{align}
    with $\delta_\epsilon \to \delta$ in the sense of distributions as $\epsilon\to0$. The functional derivative then reads
    \begin{align}
    	\frac{\delta \mathcal{Z}[J]}{\delta J(s)}
    	=
    	\lim_{\epsilon\to0}\lim_{h\to0}
    	\frac{\mathcal{Z}[J(t)+h\,\delta_\epsilon(t-s)]-\mathcal{Z}[J(t)]}{h}.
    \end{align}
    Substituting this definition back into the definition of the generating function, and assuming $\xi(t)\in C^0$, then yields
    \begin{align}
    	&\lim_{\epsilon\to0}\lim_{h\to0}
    	\int d\mu \;
    	\frac{
    	\exp\left(
    		 h\int_{|t-s|<\epsilon} dt \;\xi(t)\delta_\epsilon(t-s)
    		+\int dt\;\xi(t)J(t)
    	\right)
    	-
    	\exp\left(\int dt\;\xi(t)J(t)\right)
    	}{h} \\
    	&=
    	\lim_{\epsilon\to0}
    	\int d\mu \;
    	\left(\int_{|t-s|<\epsilon} dt \;\xi(t)\delta_\epsilon(t-s)\right)
    	\exp\left(\int dt\;\xi(t)J(t)\right) \\
    	&=
    	\int d\mu \;\xi(s)\exp\left(\int dt\;\xi(t)J(t)\right),
    \end{align}
    which is the sought-after result.
    To conclude this section, let us briefly discuss some calculation rules for the functional derivative. The Leibniz rule and linearity are maintained. Differentiation of 
	\begin{align}
    	\frac{\delta}{\delta J(s)} \int dt \; J(t)\varphi(t)
    	=\int dt \;\frac{\delta J(t)}{\delta J(s)} \varphi(t)
    	=	\varphi(s).
    \end{align}
	for some test-function $\varphi(t)$, implies that $\delta J(t) / \delta J(s)=\delta(t-s)$. Furthermore, the chain rule for some functional $F: C^1 \to C^1$ and $G: C^1 \to \mathbb{R}$ reads
	\begin{align}
		\frac{\delta G[F[J]]}{\delta J(s)} = \int dt \; \frac{\delta G[f]}{\delta f(t)} \Big|_{f=F[J(t)]} \frac{\delta F[J(t)]}{\delta J(s)}
	\end{align}
\section{Path Integral Formulation for Stochastic Differential Equations}
	To begin with, let us consider a stochastically forced differential equation of the form 
	\begin{align}
		\partial_t \phi = - F(\phi) + N(\phi) \xi(x,t) 
	\end{align}
	with a randomly distributed forcing $\xi$. We now wish to compute some observable, or statistical quantity $\mathcal{O}(\phi)$, determined through the previous evolution equation. Its expectation value can then be expressed through the probability density function $P(\xi)$, summing over all realizations of $\phi_\xi$, given some $\xi$. That is
	\begin{align}
		\langle \mathcal{O} \rangle = \int \D \xi P[\xi] \mathcal{O} (\phi_\xi)
	\end{align}
	We then employ a sifting with the infinite-dimensional delta-functional $\delta[\phi-\phi_\xi]$ to obtain.
	\begin{align}
		\langle \mathcal{O} \rangle = \int \D \xi \int \D \phi P[\xi] \mathcal{O} (\phi) \delta[\phi-\phi_\xi]
	\end{align}
	$\phi$ being a dummy integration variable and $\phi_\xi$ being the actual realization. We can then employ an infinite-dimensional change of variables to rewrite the delta functional, introducing a functional determinant
	\begin{align}
		\langle \mathcal{O} \rangle = \int \D \xi \int \D \phi P[\xi] \mathcal{O} (\phi) \delta[\partial_t \phi + F(\phi) - N(\phi) \xi(x,t)] \det \left(\partial_t + \frac{\delta F}{\delta \phi}-\frac{\delta N}{\delta \phi} \xi \right)
	\end{align}
	At this point, the subtle point arises of how one interprets the Langevin equation. That is in the sense of Ito's or Stratonovich's discretization. It can be shown  that Ito's prescription leads to a functional determinant of unity  $\mathcal{J}=\det(M)=1$ \cite{Canet2011}. 
    As a Jacobian determinant of identity significantly simplifies the following considerations, we will continue in the setting of Ito's discretization.
	At this point, we recall that the delta function can be written as
	\begin{align}
		(2 \pi)^d \delta(x-y) = \int_{\mathbb{R}^d} dp \; \exp \left(-i p (x-y) \right)
	\end{align}
	and the delta-functional, as a limit of a lattice-discretization in the region $-L \leq x \leq L$, reads as follows
	\begin{align}
		\delta[f(x)] = \lim_{N \to \infty} \prod_{k=0}^N \delta(f(Lk/N)) 
		= \lim_{N \to \infty} \prod_{k=0}^N \frac{1}{(2 \pi)^d}\left(\int_{\mathbb{R}^d} d p_k\; \exp \left(-i p_k f(Lk/N) \right)\right) \\
		=\lim_{N \to \infty} \prod_{k=0}^N \left(\frac{1}{(2 \pi)^d}\int_{\mathbb{R}} d p_k\right) \exp \left(-i \sum_{k=0}^{N} p_k f(Lk/N) \right)
		= \int \D \psi \exp \left( -i \int_{-L}^L dt \; \psi(x) f(x) \right)
	\end{align}
	finally, we may let $L \to \infty$. Inserting this definition into the equation for the observable $\mathcal{O}$ yields the following expression
	\begin{align}
		\langle \mathcal{O} \rangle = \iiint \D \phi  \D \psi  \D \xi  \exp \left(-i \int dt \; \psi \left(\partial_t \phi + F(\phi) - N(\phi) \xi(x,t)\right) \right) P[\xi]
	\end{align}
	Up until now, we did not make any assumptions about $P[\xi]$, however, it is convenient to assume a normal distribution. That is $P=1/ \sqrt{4 \pi}\exp(-1/4 \int dt \; \xi^2)$ and the observable reads
	\begin{align}
		\langle \mathcal{O} \rangle = \iiint \D \phi  \D \psi  \D \xi  \exp \left(-i \int dt \; \psi \left(\partial_t \phi + F(\phi) - N(\phi) \xi(t)\right) - \frac{1}{4}\int dt\; \xi^2 \right)
	\end{align}
	wherein the $\xi$-terms can be collected by completing the square $\left(i \psi N \xi - 1/4 \xi^2 \right)= \left(i\xi/2 + \psi N\right)^2- \psi^2 N^2$. We thus end up with a simple Gaussian integral which can be evaluated easily, resulting in an unimportant prefactor, canceled by the normalization $\langle\mathcal{O}=1\rangle=1$. The final result, making the choice $\mathcal{O} = \exp \left(\int dt \; \phi(t) J(t) \right)$ to obtain the moment-generating functional, reads
	\begin{align}\label{10}
		\mathcal{Z}[J,\overline{J}] = \iint \D \phi  \D [i\psi ]   \exp \left(-\int dt \; \psi \left(\partial_t \phi + F(\phi) + N^2 \psi \right) + \int dt \; J \phi + \overline{J} \psi \right)
	\end{align}
	\subsubsection{Ornstein-Uhlenbeck}
	As one of the simplest examples, with an analytical solution, let us briefly discuss the Ornstein-Uhlenbeck process. The governing equation and generating functional read 
	\begin{align}
		&\partial_t \phi = -\alpha \phi + \xi \\
		&\mathcal{Z} = \iint \D \phi  \D \psi \exp \left(\int dt \; -i \psi \left(\partial_t + \alpha \right) \phi - \sigma \psi^2 + J \phi \right)
	\end{align}
	With $\sigma$ the forcing's standard deviation. The action can be integrated by parts, whereafter one retrieves the definition of the delta-functional
	\begin{align}
		\mathcal{Z} &= \iint \D \phi  \D \psi \exp \left(\int dt \; -i \phi \left( \alpha - \partial_t \right) \psi - \sigma \psi^2 + J \phi \right) \\
		&=\int   \D \psi \exp \left(-\int dt  \sigma \psi^2 \right) \delta[(\alpha-\partial_t) \psi+iJ] \nonumber
	\end{align}
	The constraint $(\alpha-\partial_t) \psi+iJ=0$ can be solved easily as
	\begin{align}
		 \psi= i \int_{- \infty}^t ds \; \exp(\alpha (t-s)) J(s)
	\end{align}
	and the final expression for the generating functional turns out to be Gaussian and reads 
	\begin{align}
		\mathcal{Z} = \exp \left( \sigma \int dt  \left(\int_{- \infty}^t ds \; \exp(\alpha (t-s)) J(s) \right)^2 \right).
	\end{align}
	\section{Energy and Enstrophy Injection Rates} \label{InjectionRate}
	Later on, we will find it to be useful to consider the rate of energy or enstrophy injection as a scale-invariant quantity. Their invariance throughout the inertial subrange immediately yields insights into the scaling behaviour of correlation functions, and the critical exponents, in a fixed-point regime. The correlation function of an observable and the forcing $\xi(x,t)$ reads as follows
	\begin{align}
		\langle \xi \mathcal{O}[\phi] \rangle = \int \D \xi \; \xi  \mathcal{O}[\phi_\xi] P[\xi]= \iint \D \phi \D \xi  \; \xi \mathcal{O}[\phi] \delta[\phi_\xi-\phi] P[\xi]
	\end{align}
	we can then make a functional change of variables as before and exponentiate the delta functional to obtain
	\begin{align}
	 	\iiint \D \phi  \D \psi  \D \xi \; \xi \mathcal{O}[\phi] \exp \left(-i \int dt \; \psi \left(\partial_t \phi + F(\phi) - N(\phi) \xi(t)\right) \right) P[\xi]
	\end{align}
	Now, let us consider the following, generic, path integral for the moment
	\begin{align}
		\langle \mathcal{O}'[\xi] \rangle = \int \D \xi \mathcal{O}'[\xi]  P[\xi]
	\end{align}
	making the change of variables $\xi \to \xi + \delta \xi$ leaves the path integral invariant and we thus have the variation
	\begin{align}
		\delta \langle \mathcal{O}'[\xi] \rangle  = \int \D \xi \frac{\delta \left(\mathcal{O}'[\xi]  P[\xi] \right)}{\delta \xi (x,t)}  = \int \D \xi \left( \frac{\delta \mathcal{O}'[\xi]  }{\delta \xi (x,t)}  P[\xi] +  \frac{\delta  P[\xi] }{\delta \xi (x,t)} \mathcal{O}'[\xi]  \right) = 0
	\end{align}
	which is simply the infinite-dimensional analogue of the divergence theorem. Assuming the distribution of $\xi(x)$ to be gaussian then yields
	\begin{align}
		0 &= \int \D \xi \frac{\delta }{\delta \xi (x,t)} \left( \mathcal{O}'[\xi] \exp\left(- \frac{1}{2} \int_{x,x',t} \xi(x,t) K(x-x') \xi(x',t)\right) \right) \\
        &= 
		\delta \langle \mathcal{O}'[\xi] \rangle  = \int \D \xi \frac{\delta \mathcal{O}'[\xi]   }{\delta \xi (x,t)}  P[\xi] - \int_{x'} K(x-x') \langle \xi(x',t)  \mathcal{O}'[\xi] \rangle \nonumber
	\end{align}
	This is a version of Wick's theorem \cite{Peskin2018}. Going back to the problem at hand, and introducing the convolutional inverse $K\ast K^{-1} = \identity$ we find the following expression
	\begin{align}
		\langle \xi(t) \mathcal{O}[\phi] \rangle =  \int_{x'} K^{-1} (t-t') \langle i \psi(x',t) N(\phi(x',t))     \mathcal{O}[\phi] \rangle 		
	\end{align}
\section{The Non-Perturbative Wetterich Flow-Equation}
	Several formulations of the non-perturbative renormalization group have been derived throughout the past decades, such as the Polchinski or Cardy flow equations. However, Wetterich's flow equation \cite{Wetterich1993} is the most well-behaved and is thus the dominant formulation used today. Instead of formulating the flow in terms of $\mathcal{Z}$, one instead uses the effective action/one-particle irreducible generating functional $\Gamma$, defined by
	\begin{align} \label{11}
		\Gamma[\bm{\psi}] = \sup_{J}\Big\{ -\underbrace{\ln \left(\mathcal{Z}\right)}_{\mathcal{W}} + \int_{x} J \psi\Big\} \rightarrow 0 = \psi(x) - \frac{1}{\mathcal{Z}} \frac{\delta \mathcal{Z}}{\delta J(x)}
	\end{align}
	wherein the right-hand side follows immediately by differentiating with respect to $J(x)$. Note that within the present section we do not differentiate between purely spatial or spatiotemporal problems. We now introduce the RNG-scale $k$-dependent action and effective action, obtained by including the regulator term $ R_k$
	\begin{align}
		&S_k = S+ \int_{x,x'} \phi  R_k(x-x') \phi' \\
		&\Gamma_k[\psi] = \sup_{J_k}\Big\{ -\ln \left(\mathcal{Z}_k\right) + \int_{x} J_k \psi-\Delta S_k[\psi]\Big\}, \quad   R_k
		: \Bigg\{ \Bigg.\substack{
			\begin{aligned}
				&\lim_{k \rightarrow 0}  R_k =0 \quad  &&\lim_{k \rightarrow 0} \Gamma_k = \Gamma \\ &\lim_{k \rightarrow \Lambda}  R_k \to \infty \quad  &&\lim_{k \rightarrow \Lambda} \Gamma_k = S \label{13}
		\end{aligned}}
	\end{align}
	\begin{align} \label{14}
		\frac{\delta \Gamma_k}{\delta \psi} = J_k - \frac{\delta \Delta S_k [\psi]}{\delta \psi}
	\end{align}
	The limit $\lim_{k \rightarrow 0} \Gamma_k = \Gamma$ is trivial whereas the limit $\lim_{k \rightarrow \infty} \Gamma_k = S$ requires some attention. We first exponentiate \ref{13} and insert the Legendre relation \ref{14} to obtain
	\begin{align}
		\exp(-\Gamma_k) = \int \D \bm{\phi} \exp \left(-S[\phi]-\Delta S_k [\phi]+\Delta S_k [\psi]+ \int_{x}\left(\frac{\delta \Gamma_k}{\delta \psi}+\frac{\delta \Delta S_k[\psi]}{\delta \psi}\right)\left(\bm{\phi}-\bm{\psi}\right)\right)
	\end{align}
	we then substitute $\phi=\chi+\psi$, leaving the measure unchanged, we find
	\begin{align}
		\exp(-\Gamma_k) = \int D \chi \exp \left(-S[\psi+\chi]-\Delta S_k [\chi]+ \int_{x,t}\frac{\delta \Gamma_k}{\delta \psi} \chi\right)
	\end{align}
	For $k \to \Lambda$ the regulator term exponentially suppresses all  but the fluctuations near $\chi=0$. Whence we may employ Laplace's method to find the proposed limit.	
	Using the Legendre-relations, and introducing the correlation fucntion $\Gamma^{(2)}_k(x,y)=\delta^2 \Gamma_k / \delta \psi^i(x) \delta \psi^j(y)$, we may then find the exact Wetterich flow equation. Differentiation of equation \ref{13} yields
	\begin{align}
		\partial_k \Gamma_k &= - \frac{\partial_k \mathcal{Z}_k}{\mathcal{Z}_k} + \int_{x}\frac{1}{\mathcal{Z}_k}\frac{\delta \mathcal{Z}_k}{\delta J}\partial_k J_k + \int_{x,t} \psi \partial_k J_k - \partial_k \Delta S_k [\psi] \\
	\end{align}
	Through the Legendre relation \ref{11}, the second and third terms cancel. Differentiating the generating functional $\mathcal{Z}_k$ we find
	\begin{align}
		\frac{\partial_k \mathcal{Z}_k}{\mathcal{Z}_k} = 	\frac{1}{\mathcal{Z}_k} \int \D \bm{\phi} \partial_k \Delta S_k [\phi] \exp(...)  = \langle \partial_k \Delta S_k [\phi] \rangle
	\end{align}
	\begin{align}
		\partial_k \Gamma_k 
		&= - \frac{\partial_k \mathcal{Z}_k}{\mathcal{Z}_k} - \partial_k \Delta S_k [\psi] 
		= \int_{x,x'} \partial_k  R_k : \left(\langle \bm{\phi} \bm{\phi} \rangle - \langle \bm{\phi} \rangle \langle \bm{\phi} \rangle \right)
		=\int_{x,x'} \partial_k  R_k :  G_k
	\end{align}
	Finally, we differentiate the Legendre relation once more to obtain 
	\begin{align} \label{21}
		\frac{\delta}{\delta \psi(y)} \left(\frac{\delta \mathcal{W}_k}{\delta J(x)}-\psi(x)\right) 
		= \identity \delta(x-y)-\int_z \frac{\delta^2 \mathcal{W}_k}{\delta J(x) \delta J(z)} \frac{\delta J(z)}{\delta \psi(y)} 
		= \identity \delta(x-y)-\int_z  G_k \left[\Gamma_k^{(2)}+ R_k\right]
	\end{align}
	Combining these expressions yields the exact flow equation
	\begin{align} \label{22}
		\partial_k \Gamma_k = \mathsf{Tr} \int_{x,x'} [\Gamma^{(2)}_k+ R_k]^{-1} \partial_k R_k
	\end{align}
	Taking functional derivatives thereof then yields flow equations for the higher-order correlation functions	
	\begin{align} \label{37}
		\partial_k \Gamma^{(2)}_k(y,z) = \mathsf{Tr} \int_{x,x'} \frac{\delta^2  G_k}{\delta \psi(y) \delta \psi(z)} \partial_k  R_k	= \mathsf{Tr} \int_{x,x'}  G_k^{(2)} \partial_k  R_k	
 	\end{align}
	wherein the evaluation of $\delta^2  G_k/\delta \psi(y) \delta \psi(z)$ is a tricky subject that turns out to be resolved much easier in momentum space, at least when evaluating in a constant background field.
	\subsection{Flow Equations in Momentum/ Fourier-Space}
	The flow equations are most easily evaluated in Fourier or momentum space. In most cases, it suffices to evaluate the flow equations of some $N$-point function in a constant background field, significantly simplifying their evaluation. Observe that the we then obtain a shift invariant form of the vertex functions $\Gamma^{(2)} (x,y) = \Gamma^{(2)}(x-y)$. Taking the Fourier transform yields
	\begin{align}
		&\int_{x,y} \Gamma^{(2)}(x-y) \rvert_{\psi=\const} \exp \left(-i(x p + y q)\right) = \int_{x',y'} \Gamma^{(2)} (x') \rvert_{\psi=\const} \exp\left(-i(x'p + y' (p+q))\right) \nonumber \\
		&= (2 \pi)^d \delta(p+q) \int_{x'} \Gamma^{(2)} (x')\rvert_{\psi=\const} \exp\left(ix'p \right) 
		= (2 \pi)^d \delta(p+q) \Gamma^{(2)}\rvert_{\psi=\const} (p)
	\end{align}
	by a simple change of integration variables. Also let us drop the explicit $\rvert_{\psi=\const}$ and introduce the notation 
	\begin{align}
		\int_p = \frac{1}{(2 \pi)^d} \int_{\mathbb{R}^d} (dp)^d
	\end{align}
	with $d$ the spatial dimension. Within the following, we will use the following abuse of notation for $N$-vertex functions
    \begin{align}
        \Gamma^{(n)}\left(\{p_i\}_{i=0}^{n-1} \right) = \Gamma^{(n)}\left(\{p_i\}_{i=0}^{n-2}\right) \delta \left( \sum_{i=0}^{n-1} p_i \right)
    \end{align}
    wherein the momentum for the index/ derivative furthest to the right is eliminated. Also note the well-known convolution theorem and the following identity
	\begin{align}
		&\mathcal{F} \left[f(x) g(x)\right] = f(p) \ast g(p) = \int_q f(q) g(p-q) \\
		& \int_x f(x) = \mathcal{F}\left[f\right]_{p=0}
	\end{align}
	Inserting this into the definition of the full field-dependent propagator \ref{21} yields
	\begin{align} 
		&(2 \pi)^d \delta(p+q)\identity = \mathcal{F} \left[\int_z  G_k \left[\Gamma_k^{(2)}+ R_k\right]\right] = \left[ G_k(p,s) \ast \left[\Gamma_k^{(2)}+ R_k\right](s,q)\right]_{s=0} \label{41}\\
		&= (2 \pi)^{2d} \int_{l} \delta(p+l) \delta(q-l)  G_k (p)  \left[\Gamma_k^{(2)}+ R_k\right](p) = (2 \pi)^d \delta (p+q)  G_k (p)  \left[\Gamma_k^{(2)}+ R_k\right](p) 
		\label{42}
	\end{align}
	and thus, by factoring out $(2 \pi)^d \delta (p+q)$, reduces to a simple matrix inversion. Similarly, the Wetterich flow equation reads 
	\begin{align} 
		\partial_k \Gamma_k &= \mathsf{Tr} \left[ G_k(p,q) \ast^2 \partial_k R_k(p,q)\right]_{p=q=0} = \mathsf{Tr} \int_{s,t}   G_k(s) \delta(s+t) \ast^2 \partial_k R_k(s) \delta(-s-t) \label{37} \\
        &= \delta(0)  \mathsf{Tr} \int_{s}  G_k(s)\partial_k  R_k(s), \nonumber
	\end{align}
    or in its diagrammatic form with solid lines representing propagators and the cross symbolizing an insertion of $\partial_k R_k$
    \begin{align}
        \partial_k \Gamma_k = 
        \vcenter{\hbox{%
        \begin{tikzpicture}[scale=0.5, transform shape]
        \begin{feynman}
            \vertex [inner sep=0pt, minimum size=0pt] (a) {};
            \vertex [draw, cross out, minimum size=12pt] (b) [below=of a] {};
            \diagram*{
                (a) --[half right] (b) --[half right] (a),
            };
        \end{feynman}
        \end{tikzpicture}}}.
    \end{align}
	The prefactor of $\delta(0)$ simply results from the integration over the entire space and can be absorbed into a normalization, defining a spatial density $\Gamma/ \int_x 1$. 
	The modified flow equation for scale-dependent fields reads in momentum space
	\begin{align}
		\frac{d \Gamma_k(\psi_k)}{dk} =  \partial_k \psi_k  \Gamma_k^{(1)}(\psi_k,p=0)+  \mathsf{Tr} \int_{s}  G_k(s)\partial_k  R_k(s)	
	\end{align}
	Now, let us turn back to the flow equation \ref{37}, taking the Fourier transform thereof and invoking theorems \ref{41} and \ref{42} yields
	\begin{align} 
		\partial_k \Gamma^{(2)}_k(p,q) = \mathsf{Tr} \left[ \mathcal{F}\left[ G_k^{(2)} \right](m,n,p,q )\ast^2 \partial_k  R_k(m) \delta(m+n) \right]_{m=n=0}	
	\end{align}
	and the remaining problem turns out to be the evaluation of $\mathcal{F}\big[ G_k^{(2)}\big]$. First, let us differentiate \ref{21} with respect to $\bm{\psi}(t)$ and then take the Fourier transform thereof, dropping the explicit $\mathcal{F}[..]$'s and the renormalization-scale dependence therein,
	\begin{align}
		&0 = \int_z  G_i^{(1)}(x,z,t) \tilde{\Gamma}^{(2)}(z,y)  + G (x,z) \Gamma_i^{(3)}(z,y,t) 
		\\
        &\xrightarrow{\mathcal{F}} 0=  \left[ G_i^{(1)}(p,q,n) \ast \tilde{\Gamma}^{(2)}(q,r) +  G_k(p,q) \ast \Gamma_k^{(3)}(q,r,n)\right]_{q=0} \nonumber \nonumber \\
		& = \int_q  G_i^{(1)}(p,q,n)  \tilde{\Gamma}^{(2)}(-q,r) +  G(p,q)  \Gamma_i^{(3)}(-q,r,n) \nonumber
	\end{align}
	We may then evaluate at constant fields, gaining shift-invariance, thus
	\begin{align}
		0 &= \left(  G_i^{(1)}(p,r)   \tilde{\Gamma}^{(2)}(-r)   +  G(p) \Gamma_i^{(3)}(-r-n,r)  \right) (2 \pi)^d \delta(p+r+n) \\
        &\rightarrow   G_i^{(1)}(p,r) = -  G(p) \Gamma_i^{(3)}(p,r)   G(-r) \nonumber
	\end{align}
    \ifextra
\else
    Taking another functional derivative thereof yields the exact flow equation for the  two-point function at constant fields
    \fi
	\begin{align}\label{53}
		\partial_k \Gamma^{(2)}_{ij}(p) = \mathsf{Tr} \int_q \partial_k  R_k (q)  G (q)\left(-\Gamma_{ij}^{(4)}(q,-q,p) + 2  \Gamma_i^{(3)} (q,-p-q)    G (p+q) \Gamma_j^{(3)} (p+q,-q) \right)  G_k (q)
	\end{align}
    or expressed diagrammatically, with solid blobs representing the vertex functions $\Gamma^{(n)}$,
    \begin{align}
       \partial_k \Gamma^{(2)}_k(p)  = 2 \left(
        \vcenter{\hbox{%
        \begin{tikzpicture}[scale=0.5, transform shape]
        \begin{feynman}
            \vertex (a) {};
            \vertex[dot, minimum size=12pt, inner sep=0pt] (b) [right=of a] {};
            \vertex[draw, cross out, minimum size=12pt] (c) [below right=of b] {};
            \vertex[dot, minimum size=12pt, inner sep=0pt] (d) [above right=of c] {};
            \vertex (e) [right=of d] {};
            \diagram*{
                (a) -- (b) -- [quarter right] (c) -- [quarter right] (d) -- (e),
                (d) -- [half right] (b),
            };
        \end{feynman}
        \end{tikzpicture}}} 
        \right)
        \;- \;
        \vcenter{\hbox{%
        \begin{tikzpicture}[scale=0.5, transform shape]
        \begin{feynman}
            \vertex (a) {};
            \vertex[dot, minimum size=12pt, inner sep=0pt] (b) [below right=of a] {};
            \vertex (c) [above right=of b] {};
            \vertex[draw, cross out, minimum size=12pt, inner sep=0pt] (d) [below=of b] {};
            \diagram*{
                (a) -- (b) -- (c),
                (b) -- [half right] (d) -- [half right] (b),
            };
        \end{feynman}
        \end{tikzpicture}}}
    \end{align}
    wherein the fact was used that $G^T(p)=G(-p)$ and the cyclic property of the trace.
    \subsection{Regulator-Choice} \label{RegChoice}
    With the Wetterich flow equation at hand, let us briefly discuss the properties of the regulator matrix $\bm{R}_k$. Suppose we consider a scalar theory and the propagator has the asymptotic form $G_k \sim (p^2)^{-\alpha}$ for large momenta. Then, let's define 
    \begin{align}
        R_k \equiv \left(p^2\right)^\alpha r \left( \frac{p^2}{k^2} \right),
    \end{align}
    with $r(y)$ a dimensionless shape-function of $y=p^2/k^2$. The Wetterich flow equation then takes the form 
    \begin{align}
        \partial_k \Gamma \sim \int d^dp \frac{\partial_k r}{(1+r)} \sim \int dy y^{d/2-1} r(y)
    \end{align}
    which converges for large $y$ iff 
    \begin{align}
        \lim_{y \to \infty, \, \epsilon>0}  y^{d/2+\epsilon} r(y) = 0.
    \end{align}
    Similarly, the dominant contribution in the IR, that is, for small momenta, is 
    \begin{align}
        \lim_{\epsilon \to 0} \int_{p>\epsilon} d^d p \frac{\partial_k r}{(1+r)} \sim \lim_{\epsilon \to 0} \int_{y>\epsilon^2} y^{d/2-1} \frac{y r'}{k r} = \lim_{\epsilon \to 0} \int_{y>\epsilon^2} y^{d/2} \frac{\left( \ln{r} \right)'}{k}
    \end{align}
    which remains finite iff
    \begin{align}
        \lim_{y \to 0, \epsilon>0} y^{d/2+1+\epsilon} \left( \ln{r} \right)' = 0.
    \end{align}
    
\section{Extended Symmetries}
	The physical properties of our model equations, and the statistical flow equations derived from them, are closely related and encoded in their symmetries. 
    The following discussion on the calculation of these extended symmetries, and the notation used herein, is based upon \cite{Olver1986}, whereas the implications on the correlations functions are discussed in depth in \cite{Peskin2018}.
    Recall that every Lie group of a set of Euler-Lagrange equations $\EL(\mathcal{L})=0$ for a Lagrangian $\mathcal{L}(\bm{\phi},..,\partial^n \bm{\phi})$ yields a conserved charge with a flow carrying it, dubbed a Noether current. Let us consider an infinitesimal variational symmetry in evolutionary form
	\begin{align}
		\pr \bm{v}_Q(\mathcal{L}) = \Fd{Q}(\mathcal{L}) =\mathsf{Div}(\bm{B}) 
	\end{align}
	note also that the Euler-Lagrange equations are self-adjoint. It thus immediately follows that there is a conserved Noether charge with current $\bm{B}$. 
	To show the importance of such a transformation in the context of functional-renormalization, let us consider infinitesimal transformations leaving the functional measure $\D \bm{\phi}$ invariant and apply it to the generating functional \ref{10}
	\begin{align}
		&\mathcal{Z}[J] = \int \D\bm{\phi} \exp \left(\int_x -\mathcal{L}+ \bm{J} \bm{\phi}\right) \\
        &\xrightarrow{\bm{v}_Q} 
		0 = \delta_{\bm{v}_Q} \mathcal{Z} = \int \D\bm{\phi}   \exp \left(\int_x -\mathcal{L}-\mathsf{Div}(\bm{B})+ \bm{J} \bm{\phi} \right) \int_x \bm{J} \; \bm{v}_Q(\bm{\phi}) = \int_x \bm{J} \langle Q \rangle \nonumber
	\end{align}
	The generating functional is invariant, as it does not depend on $\bm{\psi}$ and $\bm{v}_Q$ leaves $\bm{J}$ invariant, as it is in evolutionary form.
	This can be rewritten in terms of the effective action by using the Legendre relation \ref{11} as
	\begin{align}
		0 = \int_x \frac{\delta \Gamma}{\delta \psi} \langle 	Q \rangle.
	\end{align}
	Taking functional derivatives thereof yields an infinite set of Ward-Takahashi identities. 	As a generalization of a true symmetry, a linear-contact Ward-identity is defined as the action of a Lie group which acts on the Lagrangian as follows
	\begin{align}
		\pr \bm{v}_Q(\mathcal{L}(\bm{u})) = \mathsf{Div}(\bm{B}) + \gamma(x) \bm{\phi}
	\end{align}
	again $\bm{v}_Q$ is the generator of the group action in evolutionary form. Furthermore, the variation  $\pr \bm{v}_Q(\bm{J}\bm{\phi})$ shall be linear in the dependent variables, modulo a divergence. The variation of the generating functional then reads as follows 
	\begin{align}
		\delta_{\bm{v}_Q}(\mathcal{Z})=0 = \int_x \langle \bm{\phi} \rangle \gamma + \bm{J} \langle Q \rangle 
	\end{align}
	Now, let us focus on how to find such identities. Applying the Euler operator to the invariance condition for the Lagrangian yields
	\begin{align}
		\EL (\pr \bm{v}_Q(\mathcal{L})) &= \EL (Q \cdot \EL \left(\mathcal{L}\right)) =\pr \bm{v}_Q(\EL(\mathcal{L})) + \Fd{Q}^\ast (\EL(\mathcal{L})) =\gamma(x) 
	\end{align}
	wherein $\Fd{Q}^\ast$ is the adjoint of the Frechet derivative. The previous result follows immediately from 
	\begin{align}
		\EL (\bm{A}\cdot\bm{B}) &= \Fd{A}^\ast (\bm{B}) + \Fd{B}^\ast (\bm{A})
	\end{align}
	and the fact that the Euler-Lagrange equations are self-adjoint. 
	As a necessary condition for the existence of such a group action, we may restrict ourselves to the subset $\EL(\mathcal{L})=0$, such that we obtain
	\begin{align} \label{EL82}
		\pr \bm{v}_Q(\EL(\mathcal{L})) =\gamma 
	\end{align} 
	which is easily solved using standard methods, thus giving a complete characterisation of all point- or generalized Ward-identities. Note that we additionally should restrict ourselves to characteristics $Q$, which are at most linear in the dependent variables. Applying the Euler operator to the source term of the generating functional yields
	\begin{align}
		\EL (\bm{J}\cdot \bm{Q}) &= \Fd{Q}^\ast (\bm{J})
	\end{align}
	which must be a function solely of $x$ and $\bm{J}(x)$, if we wish the total variation of $\mathcal{Z}$ to only involve the average fields and no higher-order correlation functions. Assuming a point-transformation $Q^i=(\eta^i(x,\phi)-\xi^j(x,\phi) \phi_j^i)$ yields
	\begin{align}
		\Fd{Q}^\ast(\bm{J})^i = (-D)_I \left(\frac{\partial Q^k}{\partial \phi^i_I} J^k\right) =\left(\frac{\partial \eta^k}{\partial \phi^i} - \frac{\partial \xi^l}{\partial \phi^i} \phi_l^k \right)J^k+ D_j (\xi^j J^i)
	\end{align}
	wherein kapital letters dento multi-indices and $\phi^i_J = D_J \phi^i$. the previous result can be split at the $\phi^{(1)}$-jet to yield
	\begin{align} \label{59}
		\propto \bm{\phi}^{(1)}: \; 0 = J^k \frac{\partial \xi^j}{ \partial \phi^i}-  J^i \frac{\partial \xi^j}{ \partial \phi^k} \\	
	\end{align}
	which must hold $\forall J(x)$ and thus $\xi=\xi(x)$.
	The terms dependent on $\phi$ then read
	\begin{align}
		\bm{\phi}: \; f(x,\bm{J}^{(1)}) = \left( \frac{\partial \eta^k}{\partial \phi^i} +\frac{\partial \xi^j}{\partial x^j} \delta^i_k \right)J^k + \xi^j \frac{\partial J^i}{\partial x^j}
	\end{align}
	from which it follows that $\eta$ can be at most linear in the dependent variables $\phi$.
	Now, instead at evaluating at $\EL(\mathcal{L})=0$ let us explicitly keep an arbitrary source-term and evaluate at $\EL(\mathcal{L})-\bm{J}=0$, resulting in 
	\begin{align}
		\pr \bm{v}_Q(\EL(\mathcal{L})) \rvert_{\EL(\mathcal{L})-\bm{J}=0} =\gamma-\Fd{Q}^\ast(\bm{J}) 
	\end{align} 
	We now introduce an extended generator, treating $\bm{J}$ as a dependent variable
	\begin{align}
		\bm{v}_{\tilde{Q}} = Q^i \frac{\partial}{\partial u^i} + (\zeta^i- \xi^j J_j^i) \frac{\partial}{\partial J^i} \quad \text{or equivalently} \quad
		\tilde{\bm{v}}= \bm{v}+ \zeta^i  \frac{\partial}{\partial J^i}
	\end{align}
	We  then find
	\begin{align}
		\pr \bm{v}_{\tilde{Q}}(\EL(\mathcal{L}) -\bm{J})^i\rvert_{\EL(\mathcal{L})-\bm{J}=0} = \gamma^i - J^k \left( \frac{\partial \eta^k}{\partial \phi^i} +\xi^j_j \delta^i_k \right) - \xi^j J^i_j- \left(\zeta^i- \xi^j J_j^i\right)
	\end{align}
	The monomials involving derivatives of the dependent variables vanish as before, and if we let
	\begin{align}
		\zeta^i(x,\bm{J}) = \gamma^i - J^k\left(\frac{\partial \eta^k}{\partial \phi^i}  + \xi^j_j \delta^i_k \right)
	\end{align}
	The right-hand side vanishes, and the invariance condition reads
	\begin{align}
		&\pr \bm{v}_{\tilde{Q}}(\EL(\mathcal{L}) -\bm{J})\rvert_{\EL(\mathcal{L})=\bm{J}} =0 \\
		& \tilde{\bm{v}} = \xi^i \frac{\partial}{\partial x^i} + \eta^i \frac{\partial}{\partial \phi^i} +  \left(\gamma^i - J^k\left(\frac{\partial \eta^k}{\partial \phi^i}  + \xi^j_j \delta^i_k \right)\right) \frac{\partial}{\partial J^i}
	\end{align}
	and we may find Ward identities through a Lie-group analysis of a modified set of Euler-Lagrange equations with an explicit source term and the generator as above. Note that the method outlined above is still not sufficient, and checking for invariance a posteriori remains necessary. 
	\subsubsection{Kardar Parisi Zhang}
    Before discussing ET and EIT, let us briefly investigate the properties of well studied field theories. 
	As a  first example, let us consider the Kardar-Parisi-Zhang equation in $1+1$ dimensions with action 
	\begin{align}
		S = \int_{x,t} \overline{h} \left(\partial_t h - \nu \Delta h - \frac{\lambda}{2} (\nabla h)^2 + \frac{1}{2}\overline{h} \right)
	\end{align}
	and the following modified Euler-Lagrange equations
	\begin{align}
		&\EL_{\overline{h}} = \partial_t h - \nu \Delta h - \frac{\lambda}{2} (\nabla h)^2 + \overline{h} + \gamma_{\overline{h}} \\
		&\EL_{h} =  -\partial_t \overline{h} - \nu \Delta \overline{h} + \lambda \left(\overline{h} \Delta h + \nabla h \nabla \overline{h} \right) + \gamma_h
	\end{align}
	From which we obtain the following generators, being possible candidates for generating Ward identities
	\begin{align}
		&\bm{X}_1 = \partial_t,\quad \bm{X}_2 = \partial_x \\
		&\bm{X}_3 = f(t) \partial_h, \quad \bm{X}_4 = - \lambda f(t) \partial_x + x \dot{f} \partial_h, \quad  \bm{X}_5= t \partial_t - \overline{h} \partial_{\overline{h}}
	\end{align}
	in fact, all but the last of the above generate Ward identities previously reported before. The last generator yields a scaling of $\overline{h}$ and $t$, which is not realized as a symmetry of the action, but a scaling thereof. This is a prototypical example of the non-sufficiency of the invariance condition.
	\subsubsection{Viscoelastic Burgers Equation}
	Let us now consider viscoelastic Burgers turbulence. The governing equations are given by momentum conservation
	\begin{align}
		0 = Re \, D_t u - \nabla \sigma - \beta \Delta u + \xi(x,t) 
	\end{align}
	and a constitutive equation governing the extra stress
	\begin{align}
		0 = W \Lie{u}^\alpha \sigma + g(\sigma) - 2(1-\beta) \nabla u 
	\end{align}
	wherein we introduced the parametrized objective derivative $\Lie{u}^\alpha = D_t + 2 \alpha  \, \sigma \nabla u$, such that $\alpha = 0$ corresponds to the one-dimensional corotational derivative, whereas $\alpha=\pm 1$ corresponds to the upper and lower convected Oldroyd derivative respectively. The constitutive function $g(\sigma)$ yields for $g(x)=x$ the Oldroyd-Burgers equation and for $g(x)=x^2$ yield the Giesekus-Burgers equation. The governing action thus reads
	\begin{align} \label{SBurgers}
		S^\alpha = \int_{x,t} \overline{u} \left(Re \, D_t u - \nabla \sigma -  \beta \Delta  u \right) + \overline{u}^2 + \overline{\sigma} \left(W \Lie{u}^\alpha\sigma + g(\sigma) - 2(1-\beta) \nabla u \right) 		
	\end{align}
    and we find the following Euler-Lagrange equations in the inertialess $Re=0$ limit
    \begin{align}
        &\EL_{\overline{u}} \left( \mathcal{L}^\alpha \right) = 2 \overline{u}- \nabla \sigma - \beta \Delta u \\
        &\EL_{\overline{\sigma}}\left( \mathcal{L}^\alpha \right) = W \Lie{u}^\alpha\sigma + g( \sigma ) - 2(1-\beta) \nabla u \\
        &\EL_{\sigma}\left( \mathcal{L}^\alpha \right) =- W \Lie{u}^{-\alpha} \overline{\sigma} + W \overline{\sigma} \nabla u + \frac{d g (\sigma)}{d \sigma} \overline{\sigma}+ \nabla \overline{u} \\
        &\EL_{u}\left( \mathcal{L}^\alpha \right)=- \beta \Delta \overline{u}+ 2 (\beta-1) \nabla \overline{\sigma}+ W \left( \overline{\sigma} \nabla \sigma - 2 \alpha \nabla (\sigma \overline{\sigma})\right)
    \end{align}
	Given $Re \neq 0 $, we find the following symmetries for the cases 
	\begin{align}
		\alpha \neq 0: \quad &\bm{X}_1 = \partial_t, \quad \bm{X}_2 = \partial_x, \quad \bm{X}_3 = t \partial_x + \partial_u, \quad \bm{X}_4 = \frac{t^2}{2} \partial_x + t \partial_u + \frac{Re}{2} \partial_{\overline{u}} \\
		\alpha=0, \; g=\sigma: \quad &\bm{X_5} = e^{-\frac{t}{W}} \partial_{\sigma}
	\end{align}
	whereas the following extended symmetries, that is, Ward-identity generators, emerge
	\begin{align}
		&\alpha \ne 0: \quad \bm{X}_1 = \partial_t, \quad \bm{X}_2 = f(t) \partial_x + \dot{f}(t) \partial_u, \quad \bm{X}_3 = f(t) \partial_{\overline{u}} \\
		&\alpha=0: \quad \bm{X_5} = f(t) \partial_{\sigma}
	\end{align}
	wherein we kept repeated generators that are associated with exact symmetries. Now, let us investigate the purely elastic regime with $Re=0$, where we then find the exact symmetries
	\begin{align}
		&\alpha \neq 0: \quad \bm{X}_1 = \partial_t, \quad \bm{X}_2 = f(t) \partial_x + \dot{f}(t) \partial_u,  \\
		&\alpha=0: \quad \bm{X_5} = e^{-\frac{t}{W}} \partial_{\sigma}
	\end{align}
	while the necessary condition for Ward-identity generators yields the following candidates
	\begin{align}
		&   \bm{X}_1 = \partial_t, \quad \bm{X}_2 = f(t) \partial_x + \dot{f}(t) \partial_u, \quad \bm{X}_3 = g(x,t) \partial_{\overline{u}} \\
        \alpha \neq 0: \quad & \bm{X}_5 =  \alpha W \dot{f}  x \partial_x - \alpha W f \partial_t + \alpha W \left( \ddot{f} x + 2 u \dot{f}\right) \partial_u +  \dot{f}\left(\alpha  W \sigma + \beta -1 \right) \partial_{\sigma} - 2 \alpha W  \dot{f} \overline{\sigma} \partial_{\overline{\sigma}}  \\
		\alpha=0: \quad & \bm{X_6} = f(t) \partial_{\sigma} 
	\end{align}
	The variation of the action can easily be read off directly for all generators but $\bm{X}_5$, for which we find the following variation after some manipulations
	\begin{align}
        \delta_{\bm{X}_5} \left(\mathcal{L}\right) = \left( W (2 \alpha +1) (\alpha W \sigma + \beta-1) \ddot{f}(t) + \left( (\alpha W \sigma + \beta-1) \frac{d g}{d \sigma} -2 \alpha W g\right) \dot{f}(t) \right) \overline{\sigma}.
	\end{align} 
    The quadratic part of the variation is then set to $\partial_\sigma \partial_{\overline{\sigma}}\delta_{\bm{X}_5} \left(\mathcal{L}\right)=0$. 
    Whence, the only possible material modes that admit the Ward identity $\bm{X}_5$ are the Oldroyd and Giesekus model.
    \begin{align}
        g(\sigma) &= A \sigma^2 + B \sigma \\
        f(t) &=  \exp{ \left(  \frac{(B \alpha W-2A (\beta-1)) t}{(2 \alpha +1) W^2 \alpha}\right)}, \label{X5f}
    \end{align}
    and the variation of the action thus reads as follows
    \begin{align} \label{dX5L}
        \delta_{\bm{X}_5} \left(\mathcal{L}\right) = 2 f(t) \overline{\sigma} (\beta-1)  \left(\frac{\left( A (\beta-1)-B \alpha W \right) \left( 2A (\beta-1)- B \alpha W \right) }{(2 \alpha +1) W^3 \alpha^2} \right).
    \end{align}
    \subsubsection{Elastic Turbulence in $d$-dimensions}
    The governing equations are the momentum equation and the constitutive equation, as follows
	\begin{align}
		&0 = Re\, D_t \bm{u} - \nabla \pi - \nabla \cdot \bm{\sigma} - \beta \Delta \bm{u} +\bm{\xi}(x,t)\\
		&0 = \nabla \cdot \bm{u} \\
		&0=W \Lie{\bm{u}} \bm{\sigma} + \bm{\sigma} g(\bm{\sigma}) - (1-\beta)\left(\nabla \bm{u}+ (\nabla \bm{u})^T\right)
	\end{align}
	wherein we have introduced a general isotropic function of the first stress-tensor invariants $g=g(\mathsf{Tr}(\bm{\sigma}))$ satisfying $g(0)=1$ and $\bm{\xi}$ denotes a Gaussian statistical forcing, which is itself homogeneous and isotropic as well as white-in-time.
	\begin{align}
		\langle \bm{\xi}(x,t) \bm{\xi}(x',t') \rangle = \bm{D}(x-x') \delta(t-t')
	\end{align}
	Note in particular that when choosing $g=1$, the model reduces to the  Oldroyd constitutive equation, which will be the primary focus of this work going forward.
    The Lie-derivative acts on the stress tensor and response stress as follows
    \begin{align}
        &\Lie{\bm{u}} \bm{\sigma} = D_t \bm{\sigma} + \nabla \bm{u} \cdot \bm{\sigma} + \bm{\sigma} \cdot \left(\nabla \bm{u} \right)^T \\
        &\Lie{\bm{u}} \overline{\bm{\sigma}} = D_t \overline{\bm{\sigma}}-  \overline{\bm{\sigma}} \cdot \nabla \bm{u}- \left(\nabla \bm{u} \right)^T \cdot \overline{\bm{\sigma}}  
    \end{align}
	This system of statistical differential equations is reformulated as a path integral for the generating functional $\mathcal{Z}$ with the action 
	\begin{align} \label{64}
		&S = S_M + \int_{x,t} \overline{\bm{\sigma}} : \left(W \Lie{\bm{u}}\bm{\sigma} + g \bm{\sigma} - (1-\beta)  \bm{d} \right) \\
		&S_{M} =  \int_{x,t} \overline{\bm{u}} \cdot \left(Re \, D_t \bm{u} - \nabla \pi - \nabla \cdot \bm{\sigma} - \beta \Delta \bm{u} \right) +\overline{\pi} \nabla \cdot \bm{u} + \int_{x'} \overline{\bm{u}}(x,t) \cdot \bm{D}(|x-x'|)  \cdot \overline{\bm{u}}(x',t) \nonumber
	\end{align}
    with the strain-rate $\bm{d}=\Lie{\bm{u}} \bm{g}$, where $\bm{g}$ is the metric tensor. The Euler-Lagrange equations derived from \ref{64}, assuming $g=1$ and a forcing $\propto \delta(x-x')$ for convenience, read
	\begin{align}
		& \EL_{\overline{\bm{u}}} (\mathcal{L} ) =  Re \, D_t \bm{u} - \nabla \pi -\nabla \cdot \bm{\sigma} - \beta \Delta \bm{u}  +  2 
        \overline{\bm{u}} \\
		& \EL_{\overline{\pi}}(\mathcal{L} ) = \nabla \cdot \bm{u} \\
		&\EL_{\overline{\bm{\sigma}}}(\mathcal{L} ) = W \Lie{\bm{u}}\bm{\sigma} + \bm{\sigma} - (1-\beta) \bm{d} \\
		& \EL_{\bm{u}}(\mathcal{L} ) = - Re \left(\partial_t \overline{\bm{u}}+ \bm{u} \cdot \overline{\bm{d}}\right) -\beta \Delta \overline{\bm{u}}- \nabla \overline{\pi}+ W \left(\nabla \bm{\sigma}: \overline{\bm{\sigma}}-2 \nabla \cdot( \overline{\bm{\sigma}}\cdot\bm{\sigma})\right)+2 (1-\beta) \nabla \cdot \overline{\bm{\sigma}}\\
        & \EL_{\pi}(\mathcal{L} ) = \nabla \cdot \overline{\bm{u}} \\
		& \EL_{\bm{\sigma}}(\mathcal{L} ) = -W \Lie{\bm{u}} \overline{\bm{\sigma}} + \overline{\bm{\sigma}} + \frac{1}{2} \overline{\bm{d}}
	\end{align}	
        wherein we have introduced the response strain-rate $\overline{\bm{d}}= \Lie{\overline{\bm{u}}^\sharp} \bm{g}$. Furthermore, the musical isomorphism has been applied implicitly in some places, and the equations above and the following generators should be interpreted in the usual Cartesian frame. Setting $Re=0$, the resulting Ward identities are
        \begin{align} \label{WIOb}
            & \bm{X}_1 = \partial_t, \quad \bm{X}_2 = \bm{f}(t) \cdot \partial_{\bm{x}} + \dot{\bm{f}}(t) \cdot \partial_{\bm{u}} \\
            & \bm{X}_3 =  \bm{x} \cdot \bm{\Omega} \cdot \partial_{\bm{x}} + \left( \bm{\Omega} \cdot \bm{\sigma} + \bm{\sigma}\cdot\bm{\Omega}^T\right): \partial_{\bm{\sigma}}+ \left( \bm{\Omega} \cdot \overline{\bm{\sigma}} + \overline{\bm{\sigma}}\cdot\bm{\Omega}^T\right): \partial_{\overline{\bm{\sigma}}} \quad \Omega_{ij}=f_k(t) \epsilon_{ijk} \\
            &\bm{X}_4= \bm{g}(x,t) \partial_{\overline{\bm{u}}}, \quad \bm{X}_5= g(x,t) \partial_{\pi}, \quad \bm{X}_6= g(x,t) \partial_{\overline{\pi}}
        \end{align}
        Unfortunately, all of the generators above are not particularly noteworthy. Whereas $\bm{X}_1$ corresponds to a shift in time, $\bm{X}_{2/3}$ is material frame indifference. Generators $\bm{X}_{4/5/6}$ are highly useful but can be obtained through trivial guesswork. We additionally find a scaling symmetry of the Euler-Lagrange equations, which does not carry over to the Action but induces a scaling of the Lagrange density
        \begin{align}
            &\bm{X}_7= f(t)  \bm{x} \cdot\partial_{\bm{x}} + \left( \dot{f} \bm{x} + f \bm{u} \right) \cdot\partial_{\bm{u}} - f(t) \overline{\pi} \partial_{\overline{\pi}} - f(t) \overline{\bm{\sigma}} :\partial_{\overline{\bm{\sigma}}} \\
            &\delta_{\bm{X}_7}\left(\mathcal{L}\right)= d \, f(t) \mathcal{L}
        \end{align}
        and a transformation whose variation of the action is linear in the Euler-Lagrange equations, particularly $\EL_{\overline{u}}$, which is not enforced through a Lagrange multiplier.
        The question remains if and how the previous analysis differs if one uses a material model written in terms of the corotational Jaumann derivative, as it was the case for elastic Burgulence. The Jaumann derivative \cite{Marsden2015} is defined by
        \begin{align}
            \ccd{\bm{\sigma}} &= \frac{1}{2} \left( \Lie{\bm{u}}\left( \bm{\sigma}\cdot \bm{g}^{-1} \right) \cdot\bm{g} + \bm{g}\cdot \Lie{\bm{u}}\left(\bm{g}^{-1} \cdot\bm{\sigma}  \right) \right) = D_t \bm{\sigma} +  \bm{\omega} \cdot \bm{\sigma} +  \bm{\sigma} \cdot  \bm{\omega}^T  \\
            \bm{\omega} &=   d \bm{u}^\flat \cdot \bm{g}^{-1}
        \end{align}
        that is, the our material model now corresponds to the co-rotational Jeffrey's fluid with an additional diffusive term. The corresponding action reads
        \begin{align}
            S= S_M + \int_{x,t} \overline{\bm{\sigma}} : \left(W \ccd{\bm{\sigma}} + \bm{\sigma} - (1-\beta)  \bm{d} \right)
        \end{align}
        Before discussing the variational symmetries and Ward identities thereof, let us first consider the homogeneous equations without any forcing. Aside from the usual symmetries, we find the symmetry
        \begin{align}
            \bm{X}=\bm{g} \; f\left( \tr{\bm{\sigma}} e^{\frac{t}{W}} \right) e^{-\frac{t}{W}} \partial_{\bm{\sigma}}
        \end{align}
        which results from a decoupling of the trace of $\bm{\sigma}$ from the rest of the variables. Taking the trace of the constitutive equation yields
        \begin{align}
            \tr{W \ccd{\bm{\sigma}} + \bm{\sigma} - (1-\beta)  \bm{d}} = W D_t \tr{\bm{\sigma}} + \tr{\bm{\sigma}} = 0.
        \end{align}
        Similarly, incompressibility decouples the momentum equation from the dynamics of the trace.
        This will lead to increasingly complex expressions in the variational symmetries and Ward identities, associated with a gauge freedom in $\tr{\bm{\sigma}}$. 
        Instead, it is more convenient to enforce the gauge $\tr{\bm{\sigma}}=0$.
        The Euler-Lagrange equations read
	   \begin{align}
    		& \EL_{\overline{\bm{u}}} (\mathcal{L} ) =  Re \, D_t \bm{u} - \nabla \pi -\nabla \cdot \bm{\sigma} - \beta \Delta \bm{u}  +  2 
            \overline{\bm{u}} \\
    		& \EL_{\overline{\pi}}(\mathcal{L} ) = \nabla \cdot \bm{u} \\
    		&\EL_{\overline{\bm{\sigma}}}(\mathcal{L} ) = W \ccd{\bm{\sigma}} + \bm{\sigma} - (1-\beta) \bm{d} \\
    		& \EL_{\bm{u}}(\mathcal{L} ) = - Re \left(\partial_t \overline{\bm{u}}+ \bm{u} \cdot \overline{\bm{d}}\right) -\beta \Delta \overline{\bm{u}}- \nabla \overline{\pi}+ W  \left(  \nabla \bm{\sigma} : \overline{\bm{\sigma}} +\nabla \cdot \left( \bm{\sigma}\cdot  \overline{\bm{\sigma}} -  \overline{\bm{\sigma}}\cdot \bm{\sigma} \right)\right)+2 (1-\beta) \nabla \cdot \overline{\bm{\sigma}}\\
            & \EL_{\pi}(\mathcal{L} ) = \nabla \cdot \overline{\bm{u}} \\
    		& \EL_{\bm{\sigma}}(\mathcal{L} ) = -W \ccd{\overline{\bm{\sigma}}} + \overline{\bm{\sigma}} + \frac{1}{2} \overline{\bm{d}}
	   \end{align}	
       and yields the established symmetries and Ward identities we have derived before for the Oldroyd-B fluid \eqref{WIOb}. There are no interesting new identities.
       At this point, let us briefly discuss the implications of these symmetries, or lack thereof, on the construction of closure schemes. Aside from the $\overline{u}$ sector not flowing due to $\bm{X}_5$, these are the same extended symmetries as found in the Navier Stokes setting. These can be used to determine the zero-momentum velocity sector, however, there are no implications for the stress and response stress sectors.
       Whence, nomber of degrees of freemdom that are permitted to flow is greatly increased, hindering the construction of closure schemes.

\section{Ward Identities for Elastic Burgulence}
	Let us return to the study pf $1+1$-dimensional elastic Burgulence, with the governing action \eqref{SBurgers}.
	The previously derived transformations $\bm{X}_1 = f(t) \partial_x + \dot{f}(t) \partial_u$ and $ \bm{X}_2 = g(x,t) \partial_{\overline{u}}$ then lead to the following ward identities 
	\begin{align}
		&\bm{X}_1 : \; 0 = \int_{x} \bm{J} \cdot  \nabla \langle \bm{\phi} \rangle +  \partial_t J^u \\
		&\bm{X}_2 : \; 0 = \langle 2 \overline{u} - \left(\nabla \sigma + \beta \Delta u \right) \rangle - J^{\overline{u}}
	\end{align}
	the second of which implies that the $\overline{u}$-sector of higher order vertex functions does not flow. The former can be used to determine $(\delta\Gamma^{(k)}/\delta u) (\bm{p}=0)$. Note that we choose the notation for the spatiotemporal wavenumber $p=(\bm{p}, \omega)$, similar to the notation of \cite{Peskin2018} for the four-momentum. We then find two additional identities that are specific to the models given by 
	\begin{align}
		\alpha=0, \; &A=0: \; 0 = \int_x (1-W \partial_t) \langle\overline{\sigma} \rangle - J^{\sigma} \label{X31}\\
        \alpha\neq 0: \;  &0 = \Bigg\langle\Bigg.\int_{x,t} \delta_{\bm{X}_5} \left(\mathcal{L}\right)  +W \alpha \left( \ddot{f}x+2 u \dot{f}\right) J^u + \dot{f} \left( \alpha W \sigma + \beta-1 \right) J^\sigma -2 \alpha W \dot{f} \overline{\sigma} J^{\overline{\sigma}}  \label{X32} \\
        &+ \bm{\phi} \cdot \alpha W \left( \dot{f} x \partial_x - f \partial_t \right) \bm{J} 
        \Bigg.\Bigg\rangle\nonumber 
	\end{align}
    with the variation of the action \eqref{dX5L} and the fuction $f(t)$ given by  \eqref{X5f}. 
    %
    %
    Let us denote $f(t)=\exp(\chi t)$, such that the variation reads
    \begin{align} 
        0 = \Bigg\langle \Bigg.\int_{x,t} \delta_{\bm{X}_5} \left(\mathcal{L}\right) &+ f \big( \big. (W \alpha \left( \chi^2 x+2 \chi u \right) J^u + \chi \left( \alpha W \sigma + \beta-1 \right) J^\sigma -2 \chi \alpha W \overline{\sigma} J^{\overline{\sigma}} \label{X32a1OB} \\
        &+ \bm{\phi} \cdot \alpha W \left( \chi x\partial_x - \partial_t \right) \bm{J} \big. \big) \Bigg. \Bigg\rangle \nonumber 
    \end{align}
    Let us briefly discuss the convergence of this integral. The most critical case with exponential growth for large $t$ is $A=0, B=1, \alpha =1$. For a general test function $\varphi(t) \in \mathcal{S}$, the pairing 
    \begin{align}
        \int_t \varphi(t) e^{\frac{t}{3W}} 
    \end{align}
    does not exist. However, in the present theory correlations are either instantaneous $\sim \delta^{(k)}(t)$ or originate in the constitutive equation, with solution $\sigma(x(t),t)$ along characteristics $u(x(t),t)=\dot{x}$
    \begin{align}
        \sigma(x(t),t) = \int_{-\infty}^t \exp \left( \int_{s}^t 2 \nabla u + \frac{1}{W}\right) \frac{2 (\beta-1)}{W} \nabla u (x(s),s) \; ds \sim e^{-\frac{t}{W}}.
    \end{align}
    whence, while assuming a causal response, the Ward identities associated with \eqref{X32a1OB} are expected to exist. 
    Note, that due to shift invariance we write $\mathcal{F}[f(t-t')]=f(\omega) \delta(\omega+\Omega)$. When evaluating the Fourier transform of \eqref{X32a1OB} and functional derivatives therof, we will encouter terms like $\int_t \exp(iqt+t)$, which are divergent. Nonetheless, we factor this divergent part as the ill defined $\delta(\omega-i)$, similar to the factor of $\delta(0)$ when defining the effective potential. 
    That is, we let $|T|> t$  in the bounds of integration, then factor the divergent part to find $0=(\text{some expression}) \times \int_{|t|<T} \exp(i\omega t+t)$ and set the expression in brackets to zero.
	In order to enforce the previous identity along the flow, it is convenient to choose our regulator such that the variation of $\Delta S_k$ is linear in the fields, leaving the Ward-identities for the $n>2$-point functions invariant. 
    Firstly, let us consider a scale-dependent forcing, which is promoted to a regulator,
	\begin{align}
		\int_{x',x,t} \overline{u}(x,t)  D_k (x-x') \overline{u}(x',t)
	\end{align}
	although the generating vector field $\bm{X}_5$ is divergence-free and there is no explicit $\eta^{\overline{u}}=0$, the variation of the regulator is not linear in the fields. Rather, we find the quadratic part of the variation
	\begin{align}
		\delta_{\bm{X}_5} \left[\int_{x',x,t} \overline{u} D_k (x-x') \overline{u}' \right] =  \int_{x',x,t} f(t)   \overline{u} \nabla \left((x - x') D_k \right)\overline{u}' 
	\end{align}
    This term vanishes for $D_k(x)=A \delta(x)+B/x$, with the Dirac delta term corresponding to the original white in space forcing. Additionally, we find the quadratic variation of other possible regulators
    \begin{align}
        \delta_{\bm{X}_5} \left( \Delta S_k \right)
        &\propto \int_{x,x',t}    f(t)   
            \bm{\phi} \cdot \nabla( (x-x')   \bm{R}_k (x-x')) \cdot 
        \bm{\phi}+  f(t)   
            \bm{\phi} \cdot\begin{bmatrix}
			4 & 2 & 3 & 0\\
			  & 0 & 1 & -2\\
			  &   & 2 & -1 \\
			  & & & -4 
		\end{bmatrix}
        \circ \bm{R}_k (x-x')  \cdot 
        \bm{\phi}  \nonumber
        \\
        &+ \text{linear terms} 
    \end{align}
    The condition for these breaking terms to vanish reads in momentum space as
    \begin{align} \label{RegX5}
        - \bm{p} \frac{\partial \bm{R}_k}{ \partial \bm{p}} + \bm{C} \circ\bm{R_k} = 0,
    \end{align}
    with $\circ$ denoting the Hadamard product. This can be solved explicitly as 
    \begin{align} \label{RegWI}
        R_k^{ij}(|\bm{p}|) = K_{ij}(k) \left(\frac{|\bm{p}|}{k} \right)^{C^{ij}} 
    \end{align}
    with $\bm{K}$ constants of integration that are to be chosen to make the regulator dimensionally consistent. To ensure convergence, we may only allow negative exponents in this power-law regulator.
    The propagator behaves at least as $G \sim p$, and from our previous discussion in \ref{RegChoice}, we have the condition 
    \begin{align}
        \lim_{y \to \infty, \, \epsilon>0}  y^{(d+1)/2+\epsilon} R = 0, 
    \end{align}
    which implies that all of the negative exponents are sufficient to be UV finite.
	Nonetheless, if we compute the microscale vertex function $\Gamma^{(2)}_\Lambda$, the regulator is insufficient to suppress all IR singularities of the propagator. Whence, the present choice will not be sufficient for all truncations of the Wetterich flow equation.
    \subsection{Scaling Analysis and Critical Exponents}
    We now take a closer look at the scaling behavior near the sought-after fixed point. Introducing rescaled momenta $\bm{p}\propto k$, frequencies $\omega \propto k^z$ and fields $\phi\propto k^{[\phi]}$ near the scale invariant fixed point. The Lagrangian, with $A=0$ and $B=1$ and a general forcing reads
    \begin{align}
		S^\alpha_k &= \int_{x,t}   \overline{\sigma} \left(W \Lie{u}^\alpha\sigma + \sigma - 2(1-\beta) \nabla u \right)  - \overline{u} \left( \nabla \sigma +  \beta \Delta  u \right)  + \int_{x',x,t} \overline{u} D_k (x-x') \overline{u}' \\
        &+ \text{Other Regulator Terms} \nonumber
	\end{align}
    and thus yields the canonical dimensions $[W]_c=-z$, $[\overline{u}]_c=0$, $[u]_c=z-1$, $[\overline{\sigma}]_c=1$, $[\sigma]_c=z$, $[D_k]_c=z+2$.
    Galilean invariance then enforces
    \begin{align}
        &\bm{X}_1 : \; 0 = \int_{x} \bm{J} \cdot  \nabla \langle \bm{\phi} \rangle +  \partial_t J^u \rightarrow [\bm{J} ]+[ \langle \bm{\phi} \rangle] +1 = z+ [J^u]
    \end{align}
    and thus by $[\bm{J} ]+[ \langle \bm{\phi} \rangle] = z+1$, deduced from the Legendre relations, it follows $[u]=z-1$. The shift in $\overline{u}$, with a scale-dependent forcing that is promoted to a regulator, leads to 
    \begin{align}
        \bm{X}_2 : \; &2 \int_{x'} \langle  \overline{u}'\rangle D_k = \langle\left(\nabla \sigma + \beta \Delta u \right) \rangle - J^{\overline{u}} 
        \rightarrow [\overline{u}] +[D_k] -1= [J^{\overline{u}}] =  [\sigma] +1=  [u] +2
    \end{align}
    implying the scaling $[\overline{u}] = 0$, $[\sigma]=z$ and $[D_k]=z+2$. The identities for both values of $\alpha$ first of all lead to the scaling $[W]=[t]=-z$. Depending on the case, we then find the relations
    \begin{align}
		&\alpha=0: \;[\overline{\sigma}] = [J^{\sigma}] \rightarrow  [\overline{\sigma}] = 1 \\
        &\alpha \neq 0: \;  [W]+z+1 = [J^\sigma] = [J^u]-1 = [\overline{\sigma}]  \rightarrow [\overline{\sigma}] = 1 
    \end{align}
    In order to fix the critical exponent $z$, we enforce a constant rate of energy injection, for details see \ref{InjectionRate} and dissipation across different scales. The injection rate is calculated as
    \begin{align}
        &0=\varepsilon = \langle \xi u \rangle =  \int_{x'} D_k \langle  \overline{u} u \rangle \\
        &[\varepsilon] =  [D_k] +[u] +[\overline{u}] -1 = 2 z
    \end{align}
    and we thus find the scaling $z=0$ for a scale invariant injection rate $[\varepsilon]=0$ (Note that this holds not only in the Burgers case but more generally for elastic turbulence in $d$-dimensions). 
    %
    %
    %
    As we wish the scale-dependent coupling $W_k$ to approach a non-trivial fixed point, no sub-leading corrections are possible.
    %
    With the above scaling relations in mind, let us briefly discuss the scaling of the convective operator
    \begin{align}
        \left[\int_{x,t} \nu^{-1} \, \overline{u} u \nabla u \right] = [\nu^{-1}]+[\overline{u}]+2 [u]-z \rightarrow[\nu^{-1}]= 2-z,
    \end{align}
    which is irrelevant for $z<2$. Whence, setting $Re=0$ from the outset is well justified. Note also that we only needed to utilize $\bm{X}_1$ and $\bm{X}_2$ to draw this conclusion, and the scaling is consistent not only for spatial dimension $d=1$ but for $ET$ in arbitrary $d$.
    \subsection{Mean-Field Theory and Instabilities}
    Before we delve into the application of truncations for the Wetterich equation, let us briefly discuss the stability of field configurations. The equations of motion, linearized in the vicinity of a constant field configuration with $u=0$ and no forcing, evaluated in momentum space, read
    \begin{align}
        0 &= - i \epsilon \omega u' + i \bm{p} \sigma' - \beta \bm{p}^2 u' \\
        0 &= W (i \omega \sigma' + 2 i \alpha \bm{p} u ' \sigma) + (2 A \sigma + B )\sigma ' - 2i \bm{p} (1- \beta) u',
    \end{align}
    with $\epsilon$ a regularization parameter of the kinetic term, that is, a small Reynolds number. The solution thereof yields
   \begin{align} \label{MFev}
        \omega^{\pm} &=
        \frac{
        i W \beta \bm{p}^{2}
        +i\epsilon(2A\sigma+B)
        }{2W\epsilon} \\
        & \pm \frac{
        \sqrt{
        - W^{2}\beta^{2}\bm{p}^{4}
        +4 A W \beta \epsilon \bm{p}^{2}\sigma
        +2 B W \beta \epsilon \bm{p}^{2}
        -8 \alpha W^{2}\epsilon \bm{p}^{2}\sigma
        +8(1-\beta) W\epsilon \bm{p}^{2}
        -\epsilon^{2}(2A\sigma+B)^{2}
        }}{2W\epsilon}. \nonumber
    \end{align}
    For small $\epsilon$, these two branches take the asymptotic form
    \begin{align}
        \omega^+ &\approx \frac{i \beta \bm{p}^2}{\epsilon}, \\
        \omega^- &\approx \frac{i\big((2 A \beta - 2 \alpha W)\sigma+\beta (B-2)+2\big)}{W \beta}.
    \end{align}
    The first eigenvalue is at most marginally stable for $p=0$. The latter $\omega^-$ can turn unstable depending on the mean-field value of $\sigma$. More precisely, the stable region is given by
    \begin{align}
        \sigma < -\frac{1}{2}\frac{2 + \beta (B-2)}{ \alpha W+A \beta},
    \end{align}
    for $\alpha W+A \beta>0$ and $<$ otherwise. Note that the dynamics are expected to change significantly along the renormalization flow. That is, we expect the poles of the retarded Green's $G_R$ function, as a function of the mean field, to change. For a general Martin-Siggia-Rose action, $\Gamma^{(2)}$ can be written as
    \begin{align}
        \Gamma^{(2)} = 
        \begin{bmatrix}
            M & \Gamma_A \\
            \Gamma_R & D \\
        \end{bmatrix}
    \end{align}
    evaluated at a mean field configuration with vanishing response fields $\overline{\bm{\phi}}=0$ we have 
    $M=0$, and the propagator takes the form
    \begin{align}
        G = 
        \begin{bmatrix}
            - \Gamma_R^{-1} D \Gamma_A^{-1}& \Gamma_R^{-1} \\
            \Gamma_A^{-1} & 0 \\
        \end{bmatrix},
    \end{align}
    with the ordering of the fields as $(\bm{\phi},\overline{\bm{\phi}})$, separating fields an their response counterparts.
    Whence, searching for poles of $\Gamma_R^{-1}$, the time retarded propagator, reduces to finding roots of the retarded block determinant $\det(\Gamma_R)$, that is eigenvalues $\omega(p,\sigma)$. The initial condition for the flow of $\Gamma^{(2)}$ reads
    \begin{align}
        \Gamma_{\Lambda}^{(2)}= 
        \begin{bmatrix}
			 0 & \beta \bm{p}^2 -i \epsilon \omega& i \bm{p} W \overline{\sigma} (1-2 \alpha) & -2 i\bm{p} (\alpha W\sigma + \beta -1) \\
			 & 2 & -i\bm{p} & 0\\
			  & & \overline{\sigma} g'' & g'-i \omega W \\
			 & & & 0 
		\end{bmatrix}.
    \end{align}
    and reproduces the eigenvalues derived based upon the direct linearisation of the deterministic equations \eqref{MFev}.
    \subsubsection{Leading Order Derivative Expansion}
    To gain a deeper understanding for the region of stability, and how it changes through the course graining of the RG flow, let us consider the non momentum-resolving leading order derivative expansion first. We project the renormalizaion flow onto the following ansatz for the effective action
    \begin{align} \label{GammaEff0}
        \Gamma_k= \int_{x,t} \overline{u}^2 d_k-\overline{u} \left(\nabla \sigma+ \beta \Delta u \right) + \overline{\sigma} \left(Z_k(\sigma,\overline{\sigma}) D_t \sigma + G_k (\sigma,\overline{\sigma}) + A_k(\sigma,\overline{\sigma}) \nabla u \right).
    \end{align}
    with $d_k$ the running forcing. Note that both the Galilean invariance as well as the shift in the response velocity are captured. Nonetheless, the Ward identity \eqref{X32} has not been enforced thus far. To do so is made particularly difficult due to their modification by the regulator.
    Nonetheless, we proceed with the present ansatz \eqref{GammaEff0}, and implement \eqref{X32} in its unmodified state. That is our ansatz assumes the Ward identity be restored for $k\to 0$ and that the final fixed point that is approach is not influenced by the perturbation of the trajectory through enforcement along the entire flow. Repeated functional differentiation yields the following constraints
    \begin{align} 
        Z_k &= Z_k(\rho), \quad
        A_k = X_k (\rho) (\alpha W \sigma + \beta -1),\nonumber \\
        G_k & = 
         \Sigma^2 Y_k(\rho) 
        +\Sigma \chi \frac{  
        (Z_k(\rho)+\alpha W X_k(\rho))}{  \alpha W}
        +\frac{
        \left(A(\beta-1)-B\alpha W\right)(\beta-1)}{
        \alpha^2W^2
        } \nonumber
    \end{align}
    with $\rho = \overline{\sigma}(\alpha W \sigma + \beta -1)^2 = \overline{\sigma} \Sigma^2$. Whence, the ansatz for the effective action reads
    \begin{align} \label{GammaEff1}
        \Gamma_k= \int_{x,t} \overline{u}^2 d_k -\overline{u} \left(\nabla \sigma+ \beta \Delta u \right) + \overline{\sigma} \left(Z_k(\rho) D_t \sigma + G_k(\rho,\sigma)  + X_k(\rho)  (\alpha W \sigma  + \beta-1) \nabla u \right).
    \end{align}
    Using \eqref{GammaEff1}, we obtain the following equations for the constitutive functions
    \begin{align}
        (\rho Z_k)' = -\frac{\partial}{\partial i \omega} \Gamma^{(\sigma,\overline{\sigma})} \quad
        (\rho X_k)' (\alpha W \sigma + \beta -1) = -\frac{\partial}{\partial i \bm{p}} \Gamma^{(u,\overline{\sigma})} \quad
        (\rho G_k)' =  \Gamma^{(\overline{\sigma})} 
    \end{align}
    wherein it is understood that derivatives are taken first, only then are the vertex functions evaluated at constant fields $u^\ast=\overline{u}^\ast=0$, and $\overline{\sigma}^\ast$ and zero momentum. 
    Before we delve into the governing flow equation, let us briefly discuss the implications for the mean field values implied by the ansatz. At the fixed point with $\Delta S_k=0$ and vanishing external sources $\bm{J}=0$ we find
    \begin{align}
        J^{\overline{\sigma}} &= \frac{\delta \Gamma_{k \to 0}}{\delta \overline{\sigma}} =  G_{k \to 0} + \rho G'_{k \to 0} = 0 \\
        J^\sigma &= \frac{\delta \Gamma_{k \to 0}}{\delta \sigma} = 2 \alpha W \overline{\sigma}^2 (\alpha W \sigma + \beta-1) G'_{k \to 0} = 0.
    \end{align}
    Naturally, we choose to root $\overline{\sigma}=0$ for the second equation, as usual in MSR field theory. Whence, $\rho=0$ and the first equation reduces to
    \begin{align}
        0 &= \Sigma^\ast Y_{k \to 0}(0) 
        +\chi ( \Sigma^\ast)^2 \frac{
        (Z_{k \to 0}(0)+\alpha W X_{k \to 0}(0))}{  \alpha W } 
        +\xi
    \end{align}
    with two possible solutions. For the projection point, we use the running value $\sigma^\ast(Y_k(0),X_k(0))$ that solves this equation, with field renormlization for $\overline{\sigma}$ chosen such that $Z_k(0)=W$.
    Now let us briefly discuss the stability of these configurations. The two-point function reads
    \begin{align}
        \Gamma^{(2)} &= 
       \begin{bmatrix}
			 0 & \beta \bm{p}^2 & i \bm{p} \overline{\sigma} (Z_k- \alpha W (X_k + 2\rho X_k' )) & -i \bm{p} \partial_{\overline{\sigma}} (\overline{\sigma}A_k) \\
			 & 2 d_k & -i\bm{p} & 0\\
			  & &  \overline{\sigma} \partial_\sigma^2 G_k &    \partial_\sigma \partial_{\overline{\sigma}} (\overline{\sigma}G_k) - i\omega \partial_{\overline{\sigma}} (\overline{\sigma}  Z_k) \\
			 & & &  \partial_{\overline{\sigma}}^2 (\overline{\sigma}G_k)
		\end{bmatrix} 
    \end{align}
    and evaluated at the mean field configuration yields the eigenvalues $\det (\Gamma^{(2)}_{k \to 0})=0$
    \begin{align}
        \omega = \frac{i(\alpha W \sigma^\ast + \beta-1 ) X_{k \to 0}(0)-i \beta \partial_\sigma G_{k \to 0}(0,\sigma^\ast)}{\beta W}.
    \end{align}
    The flow equations for the couplings are closed through the two-point function, which is used to compute the propagator $G$, and the following three- and four-legged vertices
    \begin{align}
        \Gamma^{(2,\overline{\sigma})}(p,q) &=
        \begin{bmatrix}
            0 & -i \bm{q} \partial_{\overline{\sigma}} (\overline{\sigma} Z_k)- i \bm{p}\partial_{\overline{\sigma}} (\overline{\sigma} \partial_\sigma A_k) & -i\bm{p} \partial_{\overline{\sigma}}^2 (\overline{\sigma} A_k) \\
            & -i(\omega+\Omega) \partial_{\overline{\sigma}} (\overline{\sigma} \partial_\sigma Z_k) + \partial_{\overline{\sigma}} (\overline{\sigma} \partial_\sigma^2 G_k) & -i \omega \partial_{\overline{\sigma}}^2 ( \overline{\sigma} Z_k ) + \partial_{\overline{\sigma}}^2 (\overline{\sigma} \partial_\sigma G_k)  \\
            &  & \partial_{\overline{\sigma}}^3 (\overline{\sigma} G_k)
        \end{bmatrix}\\
        \Gamma^{(2,\sigma)}(p,q) &=
        \begin{bmatrix}
            0 & i \bm{p} \overline{\sigma}\partial_\sigma(Z_k-\partial_\sigma A_k) & i (\bm{p}+\bm{q}) \partial_{\overline{\sigma}} (\overline{\sigma} Z_k)- i \bm{p}\partial_{\overline{\sigma}} (\overline{\sigma} \partial_\sigma A_k)\\
             & \overline{\sigma} \partial_\sigma^3 G_k & i\Omega \partial_{\overline{\sigma}} (\overline{\sigma} \partial_\sigma Z_k) + \partial_{\overline{\sigma}} (\overline{\sigma} \partial_\sigma^2 G_k)\\
             &  & i ( \omega + \Omega)\partial_{\overline{\sigma}}^2 ( \overline{\sigma} Z_k ) + \partial_{\overline{\sigma}}^2 (\overline{\sigma} \partial_\sigma G_k) 
        \end{bmatrix}\\
        \Gamma^{(2,u)}(p,q)&=
        \begin{bmatrix}
            0 & 0 & 0 \\
             & -i (\bm{p}+\bm{q}) \overline{\sigma}\partial_\sigma^2(Z_k-\partial_\sigma A_k) & -i \bm{q} \partial_{\overline{\sigma}}^2 (\overline{\sigma} Z_k)+ i (\bm{p}+\bm{q}) \partial_{\overline{\sigma}}^2 (\overline{\sigma} \partial_\sigma A_k)\\
             &  & i(\bm{p}+\bm{q}) \partial_{\overline{\sigma}}^2 (\overline{\sigma} A_k)
        \end{bmatrix}
    \end{align}
    %
  %
  \begin{align}
      \frac{\partial}{\partial i \bm{l}}\Gamma^{(2,u,\overline{\sigma})}(p,q,l) =
        \begin{bmatrix}
           0 & 0 & 0 \\
           0 &  \partial_{\overline{\sigma}} (\overline{\sigma} \partial_\sigma Z_k )- \partial_{\overline{\sigma}} (\overline{\sigma} \partial_\sigma^2 A_k ) & -  \partial_{\overline{\sigma}} (\overline{\sigma} \partial_\sigma A_k ) \\
           0 &  -  \partial_{\overline{\sigma}} (\overline{\sigma} \partial_\sigma A_k ) &- \partial_{\overline{\sigma}}^3 (\overline{\sigma} A_k)
        \end{bmatrix}
  \end{align}
  %
    %
    \begin{align}         
    \frac{\partial}{\partial i \mu}&\Gamma^{(2,\sigma,\overline{\sigma})}(p,q,l) = 
        \begin{bmatrix}
           0 & 0 & 0 \\
           0 
           & - \partial_{\overline{\sigma}} (\overline{\sigma} \partial_\sigma^2 Z_k)  & - \partial_{\overline{\sigma}}^2 (\overline{\sigma} \partial_\sigma Z_k) \\
            0& -\partial_{\overline{\sigma}}^2 (\overline{\sigma} \partial_\sigma Z_k)  & - \partial_{\overline{\sigma}}^3 ( \overline{\sigma} Z_k )
        \end{bmatrix} \nonumber
    \end{align}
    Note that we do not evaluate at $\overline{\sigma}=0$, which would also imply $\rho=0$. Using these vertices we get the following flow equations for the couplings
    \begin{align} \label{FlowG}
        \partial_k G_k = - \mathsf{Tr}\int_q \partial_k R G(q)  
        \Gamma^{(2,\overline{\sigma})}(q,-q)
        G(q) 
    \end{align}
    \begin{align}
        \partial_k (\rho Z_k)'
        &= - \mathsf{Tr}\int_q \partial_k R G(q) \Bigg\{\Bigg. \frac{\partial}{\partial i \omega} \Bigg|_{p=0} \Gamma^{(2,\sigma,\overline{\sigma})}(q,-q,p) \\
        &+2  \Gamma^{(2,\sigma)}(q,-q) G(q) \frac{\partial}{\partial i \Omega} \Gamma^{(2)}(q)  G(q) \Gamma^{(2,\overline{\sigma})}(q,-q) \nonumber \\  
        &-2\frac{\partial}{\partial i \omega} \Bigg|_{p=0}\Big( \Big.\Gamma^{(2,\sigma)}(q,-p-q) G(q)  \Gamma^{(2,\overline{\sigma})}(p+q,-q)        \Big.\Big)\Bigg.\Bigg\} G(q)  \nonumber        
    \end{align}
    \begin{align}
        \partial_k (\rho X_k)' &= -\frac{1}{\Sigma}  \mathsf{Tr}\int_q \partial_k R G(q) \Bigg\{\Bigg. \frac{\partial}{\partial i \bm{p}} \Bigg|_{p=0} \Gamma^{(2,u,\overline{\sigma})}(q,-q,p) \\
        &+2  \Gamma^{(2,u)}(q,-q) G(q) \frac{\partial}{\partial i \bm{q}} ( \Gamma^{(2)}(q) + R_k(q))  G(q) \Gamma^{(2,\overline{\sigma})}(q,-q) \nonumber \nonumber \\  
        &-2\frac{\partial}{\partial i \bm{p}} \Bigg|_{p=0}\Big( \Big.\Gamma^{(2,u)}(q,-p-q) G(q)  \Gamma^{(2,\overline{\sigma})}(p+q,-q)            
        \Big.\Big)\Bigg.\Bigg\} G(q) \nonumber         
    \end{align}
    All of which are implicitly evaluated at the running $\sigma^\ast$, or equivalently $\Sigma^\ast$. The derivatives with respect to the new variables read
    \begin{align}
        \frac{\partial}{\partial \sigma} &= \alpha W \left( \frac{\partial}{\partial \Sigma} + \frac{2 \rho}{\Sigma} \frac{\partial}{\partial \rho}\right) \\
        \frac{\partial}{\partial \overline{\sigma}}& = \Sigma^2 \frac{\partial}{\partial \rho}
    \end{align}
    Now, let us introduce dimensionless variables, denoted by hats
    \begin{align}
        f (k,\rho=\hat{\rho} \exp ([\rho]\ln{k})) &= k^{[f]} \hat{f}(k,\hat{\rho}) \\
        k \partial_k f(k,\rho)+\rho [\rho] \partial_{\rho} f(k,\rho) &= k^{[f]} ([f] \hat{f} + k\partial_k \hat{f}) \\
        k\partial_k \hat{f} (k, \hat{\rho}) &= k^{1-[f]} \partial_k f (k,\rho)+ [\rho] \partial_{\hat{\rho}} \hat{f}-[f] \hat{f}
    \end{align}
	\subsection{Momentum Resolving Closure}
    With the extended symmetries of the Lagrangian at hand, we may also derive a closure of the two-point equation in the spirit of the Blaizot Mendez Wschebor (BMW) scheme \cite{Blaizot2006}. In stark contrast to Navier-Stokes turbulence \cite{Canet2022} or the Kardar-Parisi-Zhang equation \cite{Canet2011a}, we do not gain access to the full zero-momentum sector of $\Gamma^{(k)}$ but only to the velocity and response-velocity sector thereof. First and foremost, the shift in $\overline{u}$ yields
    \begin{align}
		0 = \Gamma^{(\overline{u},k >1)} 
	\end{align}
    similar to the pressure term in Navier-Stokes turbulence which does not flow. Furthermore, material frame indifference yields 
	\begin{align}
		&0 = i \omega \Gamma^{(u,2)}_{ij} \left(\substack{0 \\ \omega},\substack{\bm{q} \\ \Omega}\right) - i \bm{q} \left(\Gamma^{(2)}_{ij}\left(\substack{\bm{q} \\ \Omega+\omega}\right) - \Gamma^{(2)}_{ij}\left(\substack{\bm{q} \\ \Omega}\right)   \right) \\
		& 0 =  i\omega \Gamma^{(u,3)}_{ijk} \left(\substack{0 \\ \omega}, \substack{\bm{q} \\ \Omega},\substack{\bm{l} \\ \mu}\right) - i\left( \bm{q} \Gamma^{(3)}_{ijk} \left(\substack{\bm{q} \\ \Omega+\omega},\substack{\bm{l} \\ \mu}\right)+  \bm{l}\Gamma^{(3)}_{ijk} \left(\substack{\bm{q} \\ \Omega},\substack{\bm{l} \\ \mu+\omega}\right) -  (\bm{l} + \bm{q} )\Gamma^{(3)}_{ijk} \left(\substack{\bm{q} \\ \Omega},\substack{\bm{l} \\ \mu}\right)\right)
	\end{align}
    and we may thus set $u=0$ and close the velocity sector as $\Gamma^{(u,2)}\left(\substack{\bm{p} \\ \omega},\substack{\bm{q} \\ \Omega}\right) \approx \Gamma^{(u,2)}_{ij} \left(\substack{0 \\ \omega},\substack{\bm{q} \\ \Omega}\right)$ and similar for the four-vertex function with
    \begin{align}
        &\Gamma^{(u,2)} \left(\substack{0 \\ \omega},\substack{\bm{q} \\ \Omega}\right) = \frac{\bm{q}}{\omega}\left(\Gamma^{(2)}_{ij}\left(\substack{\bm{q} \\ \Omega+\omega}\right) - \Gamma^{(2)}_{ij}\left(\substack{\bm{q} \\ \Omega}\right)   \right) = \mathscrsfs{A}_1(\omega) \left[\Gamma^{(2)}\left(\substack{\bm{q} \\ \Omega}\right) \right] \\
        &\Gamma^{(u,3)} \left(\substack{0 \\ \omega}, \substack{\bm{q} \\ \Omega},\substack{\bm{l} \\ \mu}\right) =  \frac{1}{\omega}\left( \bm{q} \Gamma^{(3)} \left(\substack{\bm{q} \\ \Omega+\omega},\substack{\bm{l} \\ \mu}\right)+  \bm{l}\Gamma^{(3)} \left(\substack{\bm{q} \\ \Omega},\substack{\bm{l} \\ \mu+\omega}\right) -  (\bm{l} + \bm{q} )\Gamma^{(3)} \left(\substack{\bm{q} \\ \Omega},\substack{\bm{l} \\ \mu}\right)\right) =\mathscrsfs{A}_2(\omega) \Gamma^{(3)} \left( \substack{\bm{q} \\ \Omega},\substack{\bm{l} \\ \mu}\right) \\
        &\Gamma^{(u,u,2)} \left(\substack{0 \\ \omega},\substack{0 \\ \Omega},\substack{\bm{l} \\ \mu}\right) = \mathscrsfs{A}_1(\Omega) \mathscrsfs{A}_1(\omega) \left[\Gamma^{(2)}\left(\substack{\bm{l} \\ \mu}\right) \right]
    \end{align}
    However, we may do considerably better by utilizing the Ward identity \ref{X32}, generated by $\bm{X}_5$. First, however, recall the definition of $\chi$ 
    \begin{align}
        \chi = \frac{B \alpha W-2A (\beta-1)}{(2 \alpha +1) W^2 \alpha}.
    \end{align}
    Differentiation of \ref{X32} yields
	%
    %
    \begin{align}
		\propto \delta': \; &0 = i \Gamma^{(u,2)}_{ij} \left(\substack{0 \\ i \chi},\substack{\bm{q} \\ \Omega}\right) - \frac{\bm{q}}{\chi}\left( \Gamma^{(2)}_{ij} \left(\substack{\bm{q} \\ \Omega+i \chi}\right)-\Gamma^{(2)}_{ij} \left(\substack{\bm{q} \\ \Omega}\right)\right) \\
		\propto \delta: \; &0 = \big( \alpha\chi  W \left(2 u + i \chi \partial_{\bm{p}}\right) \Gamma^{(u,2)}_{ij} +\chi (\alpha W \sigma + \beta -1) \Gamma^{(\sigma,2)}_{ij}-2 \alpha \chi W \overline{\sigma} \Gamma^{(\overline{\sigma},2)}_{ij}\big)\left(\substack{0 \\ i\chi},\substack{\bm{q} \\ \Omega}\right) \\
        &- \alpha i \Omega W \left(  \Gamma^{(2)}_{ij} \left(\substack{\bm{q} \\ \Omega+i\chi}\right)-   \Gamma^{(2)}_{ij} \left(\substack{\bm{q} \\ \Omega}\right) \right) - \alpha  \chi W \bm{q}  \partial_{\bm{q}} \Gamma^{(2)}_{ij} \left(\substack{\bm{q} \\ \Omega+i \chi}\right) \nonumber \\
		&+2\alpha \chi W  \left(\delta^i_u \Gamma^{(u,1)}_{j} \left(\substack{\bm{q} \\ \Omega+i\chi}\right)  +\delta^j_u \Gamma^{(1,u)}_{i}\left(\substack{\bm{q} \\ \Omega}\right)-\delta^i_{\overline{\sigma}} \Gamma^{(\overline{\sigma},1)}_{j} \left(\substack{\bm{q} \\ \Omega+i \chi}\right) -\delta^j_{\overline{\sigma}} \Gamma^{(1,\overline{\sigma})}_{i} \left(\substack{\bm{q} \\ \Omega}\right) \right) \nonumber\\
		& + \alpha \chi W \left(\delta^i_\sigma \Gamma^{(\sigma,1)}_{j} \left(\substack{\bm{q} \\ \Omega+i\chi}\right) +\delta^j_\sigma \Gamma^{(1,\sigma)}_{i} \left(\substack{\bm{q} \\ \Omega}\right) \right)  \nonumber
	\end{align}
	The terms $\propto \delta'$ then vanish due to material frame indifference and analytic continuation in the upper half $\omega$-plane for positive $\chi$. 
	This Ward-identity then suggests a closure of the three-vertex term in the Wetterich flow equation as follows
    \begin{align}
		\Gamma^{(u,2)} (p,q) \approx \Gamma^{(u,2)} \left(\substack{0 \\ \omega},\substack{\bm{q} \\ \Omega}\right) + \bm{p} \, \partial_{\bm{p}}\Gamma^{(u,2)} \left(\substack{0 \\ i \chi},\substack{\bm{q} \\ \Omega}\right)
	\end{align}
    However, it remains to evaluate the $\sigma$ and $\overline{\sigma}$ sectors in the previous identity. Assuming $|\Omega|\gg |\chi|$ suggests replacing the vertices evaluated at zero momentum and thus
	\begin{align}
		&\partial_{\bm{p}} \Gamma^{(u,2)}_{ij}\left(\substack{0 \\ i \chi},\substack{\bm{q} \\ \Omega}\right) \approx i \left( \frac{\alpha W \sigma + \beta -1}{\alpha \chi W}\partial_{\sigma}-\frac{2 \overline{\sigma}}{\chi}  \partial_{\overline{\sigma}} \right) \Gamma^{(2)}_{ij}\left(\substack{\bm{q} \\ \Omega}\right) \\
        &+ \frac{\Omega}{\chi^2} \left(\Gamma^{(2)}_{ij} \left(\substack{\bm{q} \\ \Omega+i\chi}\right)-   \Gamma^{(2)}_{ij} \left(\substack{\bm{q} \\ \Omega}\right) \right) 
        -  \frac{\bm{iq}}{\chi}  \partial_{\bm{q}} \Gamma^{(2)}_{ij} \left(\substack{\bm{q} \\ \Omega+i/(3W)}\right) \nonumber  \\
		&+ \frac{2i }{\chi} \left(\delta^i_u \Gamma^{(u,1)}_{j} \left(\substack{\bm{q} \\ \Omega+i/ (3W)}\right)  +\delta^j_u \Gamma^{(1,u)}_{i}\left(\substack{\bm{q} \\ \Omega}\right)-\delta^i_{\overline{\sigma}} \Gamma^{(\overline{\sigma},1)}_{j} \left(\substack{\bm{q} \\ \Omega+i\chi}\right) -\delta^j_{\overline{\sigma}} \Gamma^{(1,\overline{\sigma})}_{i} \left(\substack{\bm{q} \\ \Omega}\right) \right) \nonumber \\
		& + \frac{i}{\chi} \left(\delta^i_\sigma \Gamma^{(\sigma,1)}_{j} \left(\substack{\bm{q} \\ \Omega+i\chi}\right) +\delta^j_\sigma \Gamma^{(1,\sigma)}_{i} \left(\substack{\bm{q} \\ \Omega}\right) \right)  
        =\mathscrsfs{B}_1 \left[\Gamma^{(2)} \left(\substack{\bm{q} \\ \Omega}\right)\right]  \nonumber
	\end{align}
	Equivalently, for the four-point functions, we find the following identity
	\begin{align}
		\propto \delta': \; & 0 = i \Gamma^{(u,3)}_{ijk}\left(\substack{0 \\ i \chi},\substack{\bm{q} \\ \Omega},\substack{\bm{l} \\ \mu}\right) +\frac{1}{\chi} \left(\bm{q} \Gamma^{(3)}_{ijk} \left(\substack{\bm{q} \\ \Omega+i\chi},\substack{\bm{l} \\ \mu }\right) +  \bm{l} \Gamma^{(3)}_{ijk} \left(\substack{\bm{q} \\ \Omega},\substack{\bm{l} \\ \mu +i\chi }\right)  - \left(\bm{l}-\bm{q}\right) \Gamma^{(3)}_{ijk}\left(\substack{\bm{q} \\ \Omega},\substack{\bm{l} \\ \mu }\right) \right)  \\
		\propto \delta: \; & 0 = \big( \alpha \chi W \left(2u+i \chi \partial_{\bm{p}} \right)\Gamma^{(u,3)}_{ijk} + \chi (\alpha W \sigma + \beta -1) \Gamma^{(\sigma,3)}_{ijk}-2 \alpha \chi W \overline{\sigma} \Gamma^{(\overline{\sigma},3)}_{ijk}\big)\left(\substack{0 \\ i \chi},\substack{\bm{q} \\ \Omega},\substack{\bm{l} \\ \mu}\right) \\
        &- \alpha \chi W \left(\bm{q} \partial_{\bm{q}} \Gamma^{(3)}_{ijk} \left(\substack{\bm{q} \\ \Omega+i\chi},\substack{\bm{l} \\ \mu }\right)  + \bm{l}   \partial_{\bm{l}} \Gamma^{(3)}_{ijk} \left(\substack{\bm{q} \\ \Omega},\substack{\bm{l} \\ \mu +i\chi } \right)  \right) \nonumber \\
		&- \alpha i W \left( \Omega \Gamma^{(3)}_{ijk} \left(\substack{\bm{q} \\ \Omega+i\chi},\substack{\bm{l} \\ \mu }\right) + \mu \Gamma^{(3)}_{ijk} \left(\substack{\bm{q} \\ \Omega},\substack{\bm{l} \\ \mu +i\chi}\right) - (\Omega+\mu)\Gamma^{(3)}_{ijk} \left(\substack{\bm{q} \\ \Omega},\substack{\bm{l} \\ \mu }\right) \right) \nonumber \\
		&-2 \alpha \chi W \left(\delta^i_{\overline{\sigma}} 	\Gamma^{(\overline{\sigma},2)}_{jk} \left(\substack{\bm{q} \\ \Omega+i\chi},\substack{\bm{l} \\ \mu}\right) +\delta^j_{\overline{\sigma}} \Gamma^{(1,\overline{\sigma},1)}_{ik} \left(\substack{\bm{q} \\ \Omega},\substack{\bm{l} \\ \mu + i\chi}\right) +\delta^k_{\overline{\sigma}} \Gamma^{(2,\overline{\sigma})}_{ij} \left(\substack{\bm{q} \\ \Omega},\substack{\bm{l} \\ \mu }\right)  \right) \nonumber \\
		&+2 \alpha \chi W \left(\delta^i_u \Gamma^{(u,2)}_{jk} \left(\substack{\bm{q} \\ \Omega+i\chi},\substack{\bm{l} \\ \mu}\right)  +\delta^j_u \Gamma^{(1,u,1)}_{ik}\left(\substack{\bm{q} \\ \Omega},\substack{\bm{l} \\ \mu + i\chi}\right) +\delta^k_u \Gamma^{(2,u)}_{ij}\left(\substack{\bm{q} \\ \Omega},\substack{\bm{l} \\ \mu }\right)  \right) \nonumber \\
		& + \alpha \chi W \left(\delta^i_{\sigma} 	\Gamma^{(\sigma,2)}_{jk} \left(\substack{\bm{q} \\ \Omega+i\chi},\substack{\bm{l} \\ \mu}\right) +\delta^j_{\sigma} \Gamma^{(1,\sigma,1)}_{ik} \left(\substack{\bm{q} \\ \Omega},\substack{\bm{l} \\ \mu + i\chi}\right) +\delta^k_{\sigma} \Gamma^{(2,\sigma)}_{ij} \left(\substack{\bm{q} \\ \Omega},\substack{\bm{l} \\ \mu }\right)  \right) \nonumber   		
	\end{align}
    \begin{align}
        \partial_{\bm{p}} \Gamma^{(u,3)} \left(\substack{0 \\ i \chi},\substack{\bm{q} \\ \Omega},\substack{\bm{l} \\ \mu}\right) = \mathscrsfs{B}_2 \left[\Gamma^{(3)} \left(\substack{\bm{q} \\ \Omega},\substack{\bm{l} \\ \mu}\right)\right]
    \end{align}
    wherein the term $\propto \delta'$ vanishes as before. The unclosed vertices therein involving $\sigma$ and $\overline{\sigma}$ sectors are again closed by setting their momenta to zero. This identity can then be used to obtain a closure similar as for the three-point functions
    \begin{align}
		\Gamma^{(u,u,2)} (p,q,l) &\approx \Gamma^{(u,u,2)} \left(\substack{0 \\ \omega},\substack{0 \\ \Omega},\substack{\bm{l} \\ \mu }\right) + \bm{p} \partial_{\bm{p}}  \Gamma^{(u,u,2)} \left(\substack{0 \\ i\chi},\substack{0 \\ \Omega},\substack{\bm{l} \\ \mu }\right)
		+ \bm{q} \partial_{\bm{q}}  \Gamma^{(u,u,2)} \left(\substack{0 \\ \omega},\substack{0 \\ i\chi},\substack{\bm{l} \\ \mu }\right) \\
		&+ \bm{p} \bm{q} \partial_{\bm{p}} \partial_{\bm{q}}  \Gamma^{(u,u,2)} \left(\substack{0 \\ i\chi},\substack{0 \\ i\chi},\substack{\bm{l} \\ \mu }\right) \nonumber
    \end{align}
    \ifextra
\else
    Although the BMW closure is a natural choice when implementing the Ward identity \eqref{X32a1OB}, several obstructions remain. Mainly, the fact that is breaks a general regulator, leading to modification terms of the identities. Although these difficulties can be overcome, they lead to significant numerical difficulties.
    One is either forced to solve a nonlinear system of equations for the vertex functions at each RG timestep, or one has to utilize a lower order truncation in the vertices appearing within the induced Ward identity modification terms.
    Hence the evaluation of the closure is outside the scope of the present work and will be left to future publications. 
    \fi

\section{Conclusion and Outlook}

In this work, we formulated elastic and elasto-inertial turbulence within the Martin-Siggia-Rose formalism and investigated the resulting field theory by means of a symmetry analysis in preparation for an application of the functional renormalization group. In particular, we developed a systematic method for constructing Ward identities directly from the Euler-Lagrange equations with source terms through an extension of standard methods. This yields an algorithmic route to both exact symmetries and linear-contact Ward identities, and thereby provides a practical way of extracting nonperturbative constraints before introducing any truncation.

Applied to viscoelastic turbulence, this analysis shows in a precise sense why the functional renormalization group is more difficult here than in the Navier-Stokes case. In Navier-Stokes turbulence, Galilean invariance and the associated Ward identities strongly constrain the zero-momentum sector and provide substantial structure for closure schemes. In contrast, for elastic and elasto-inertial turbulence the additional stress sector carries far fewer protecting symmetries. While the velocity sector still inherits the familiar constraints associated with translations and response-field shifts, the stress and response-stress sectors remain comparatively unconstrained. Moreover, there is in general no symmetry enforcing a vanishing mean stress. As a consequence, one must expect a larger running theory space, a nontrivial mean-field background, and the possible appearance of dynamical instabilities already at the level of the effective action.

These features are already visible in the dimensionally reduced model of elastic Burgulence studied here. The Burgers setting retains the essential coupling between velocity and elastic stress while remaining simple enough to analyse in detail. The Ward identities obtained in this model constrain the admissible form of the effective action and directly motivate symmetry-adapted truncation schemes. In particular, the response-velocity sector is protected from renormalization, while the remaining identities severely restrict the functional dependence of the constitutive couplings in a derivative expansion. This makes Burgulence a useful intermediate model as it is rich enough to display the difficulties specific to viscoelastic turbulence, but simple enough to test approximation strategies in a controlled way.

On this basis, we outlined two complementary closure strategies. The first is a symmetry-adapted leading-order derivative expansion, designed to preserve the available Ward identities and to follow the renormalization of the constitutive sector together with the mean-field stability properties. The second is a momentum-resolving closure formulated on a lattice in momentum space, which is closer in spirit to Blaizot-Mendez-Wschebor-type schemes and is intended to retain more of the nonlocal momentum dependence of the flow. The former is better suited for extraction of the large-scale physics, in particular phase diagrams encoding the region of stability, whereas the latter resolves the momentum dependence. Taken together, these schemes provide a concrete route toward nonperturbative approximations for elastic Burgulence and, ultimately, for elastic turbulence itself.

The principal conclusion is therefore twofold. First, the symmetry content of elastic and elasto-inertial turbulence is substantially weaker than in Navier-Stokes turbulence, and this difference has direct consequences for the feasibility and design of closures. Second, the source-extended Lie-group method developed here provides a practical algorithm for deriving the Ward identities needed to build such closures systematically rather than heuristically, by gauging existing symmetries. In that sense, the present work establishes the structural foundation for subsequent quantitative calculations.

 \pagebreak

\printbibliography[resetnumbers, heading=none]

\end{document}